\documentclass[english]{article}
\usepackage{mathpazo}
\usepackage[T1]{fontenc}
\usepackage[utf8]{inputenc}
\usepackage{babel}
\usepackage{array}
\usepackage{booktabs}
\usepackage{multirow}
\usepackage{amsmath}
\usepackage{amstext}
\usepackage{amsthm}
\usepackage{graphicx}
\usepackage{threeparttable}
\usepackage[dvipsnames]{xcolor}
\usepackage{rotating}

\usepackage[normalem]{ulem} 
\usepackage{titling} 
\usepackage{lineno} 

\usepackage[authoryear]{natbib}
\usepackage{hyperref}
\usepackage{extraipa}
\usepackage{setspace}
\usepackage{bbm}
\usepackage{fullpage}
\usepackage{subfig}
\usepackage{mathtools}

\hypersetup{unicode=true,
bookmarksnumbered=false,bookmarksopen=false,
 breaklinks=false,pdfborder={0 0 0},pdfborderstyle={},colorlinks=false}

\makeatletter

\providecommand{\tabularnewline}{\\}

\newcommand{\mycomp}[2]{{#1}{\scriptstyle\div}{#2}}

\DeclareMathOperator{\E}{E}
\DeclareMathOperator{\var}{Var}
\DeclareMathOperator{\Var}{Var}
\DeclareMathOperator{\Cov}{Cov}

\theoremstyle{plain}
\newtheorem{thm}{\protect\theoremname}
\newtheorem{prop}[thm]{\protect\propositionname}

\providecommand{\propositionname}{Proposition}
\providecommand{\theoremname}{Theorem}

\newcommand{\tablesource}[1]{
{\setstretch{1}
\raggedright{\small
Source: #1}
\par}}

\newcommand{\tablecomment}[1]{
{\setstretch{1}
\raggedright{\small #1}
\par}}

\@ifundefined{showcaptionsetup}{}{%
 \PassOptionsToPackage{caption=false}{subfig}}

\makeatother

\begin{document}
\title{ On the benefits of index insurance in US agriculture:\\ a large-scale analysis using satellite data}
\author{Matthieu Stigler \thanks{Corresponding author, \url{Matthieu.Stigler@gmail.com}. We thank participants at the AAEA Summer Meeting, UC Davis Big Ag Data Conference, and the Agricultural Policy Conference 2020 for useful comments. We thank also Michael Carter, Mario Miranda, Elinor Benami, Andrew Hobbs, Jon Einar Flatnes, Jisang Yu, Zara Khan and Sylvain Coutu for helpful feedback and comments.}, David Lobell\\
Stanford University, Center for Food Security and the Environment}

\date{ January 29 2021 (latest version 
\textcolor{blue}{\normalsize{}\uline{\href{https://arxiv.org/abs/2011.12544}{here}}})}

\maketitle


\begin{abstract}
Index insurance has been promoted as a promising solution for reducing agricultural risk compared to traditional farm-based insurance. By linking payouts to a regional factor instead of individual loss, index insurance reduces monitoring costs, and alleviates the problems of moral hazard and adverse selection. Despite its theoretical appeal, demand for index insurance has remained low in many developing countries, triggering a debate on the causes of the low uptake. Surprisingly, there has been little discussion in this debate about the experience in the United States. The US is a unique case as both farm-based and index-based products have been available for more than two decades. Furthermore, it features a large number of insurance zones, allowing interesting comparisons over space. 


In this paper, we investigate the benefits and determinants of index insurance in the US leveraging a field-level dataset for corn and soybean obtained from satellite predictions. While previous studies were based either on county aggregates or on relatively small farm-level dataset, our satellite-derived data gives us a very large number of fields (close to 1.8 million) comprised within a large number of index zones (600) observed over 20 years. We run a large-scale simulation comparing the benefits of each insurance scheme. We make two main contributions. First, we show that in our simulations, demand for index insurance is unexpectedly high, at about 30\% to 40\% of total demand. This result is robust to relaxing several assumptions of the model and to using prospect theory instead of expected utility. Second, we examine the spatial determinants of the suitability of index insurance across 600 counties. Our results indicate that the choice of metric to assess the suitability of insurance can lead to opposite results. When assessed against no insurance, index insurance is most beneficial in the counties with highest temporal variability. When assessed against farm-level insurance instead, index insurance is now the least beneficial in those same counties. Taken together, our results contribute towards improved policy design by shedding a more optimistic light on the overall usefulness of index insurance, and by deepening our understanding of the spatial factors constraining its spread.



\end{abstract}

\vspace{1cm}

\section{Introduction}

Risk is ubiquitous in agriculture. Weather has an important influence on production, yet remains difficult to predict. Likewise, agricultural prices are typically very volatile, as experienced for example during the price spike in 2007-2008. This risk has several negative consequences on farmers. In presence of risk, farmers reduce output, and opt for low-yielding low-risk technologies \citep{Dercon1998,Dercon2002}. Furthermore, in developing countries with missing credit markets, risk affects negatively farmer's ability to smooth consumption and reduces both demand and supply of credit \citep{BoucherCarterEtAl2008,KarlanOseiEtAl2014}. 

Agricultural insurance is an important tool to reduce the risk faced by farmers. Historically, initial insurance instruments focused on indemnity-based schemes, where payouts are triggered when yields on a given field fall below a certain percentage of the field's expected yield. However, this \emph{field-based} scheme suffers from multiple issues: 1) moral hazard, where being insured leads to taking undue risk, 2) adverse selection, where the possible under and over-evaluation of individual
risk leads to adverse sorting of farmers, and finally 3) high monitoring costs due to the requirement of assessing damage and the data needed for pricing individual premiums. As a response to these issues, index insurance offers an interesting alternative. Index insurance links  the insurance payout to low realizations of an external index, which is often defined based on output (average yields in a given area) or inputs (weather variables such as rainfall, temperature, etc). The advantages of index insurance are reduced costs as monitoring individual fields is no longer necessary, absence of moral hazard since farmers individual actions have no influence on the index, and potentially reduced adverse selection.\footnote{Note that adverse selection due to spatial or temporal variations in the accuracy of the index is still possible, see \citet{JensenMudeEtAl2018}.} These advantages of index insurance over traditional indemnity-based insurance have led to the implementation of several schemes throughout the world, in particular in developing countries, and to a sustained interest in the literature \citep[see the reviews by][]{BarnettMahul2007,MirandaFarrin2012,CarterJanvryEtAl2017}. 

Despite the theoretical appeal of index insurance, success of the various schemes implemented is rather limited, as summarized by \citet{Binswanger-Mkhize2012} provocative title, \emph{Is There Too Much Hype about Index-based Agricultural Insurance?} In general, take-up is found to be very low, even at subsidized premium rates, questioning the sustainability of such schemes without public subsidies (\citealp{ColeGineEtAl2013,ColeXiong2017}). The main culprit lies in the principle itself of index insurance: by de-linking payouts from individual losses, index insurance introduces \emph{basis risk}, i.e. the probability that a farmer experiences a loss whereas the index does not lead to a payout. Ultimately, basis risk is a function of the index accuracy, and hence depends on whether aggregate yields (for outcome-based indices) or specific rainfall variables (for input-based indices) predict well individual yields. While basis risk is widely acknowledged as the main issue of index insurance, few studies yet have been able to measure it in practice. Among the few of those, \citet{JensenBarrettEtAl2016} analyze a livestock index insurance program in Kenya using four years of data, and conclude with a cautionary note about index insurance, finding a substantial basis risk. 

In this paper, we take advantage of satellite data techniques to construct a very large dataset of field-level yields for corn and soybeans in the Corn Belt area of the United States of America. 
Using satellite data allows us to observe almost every field within a large zone, which has the double advantage of permitting both a fine-scale analysis within a given zone as well as a large-scope analysis comparing many insurance zones. Focusing on the US Corn Belt offers an interesting case study for two reasons.
Firstly, its large and rather uniform fields offer a particularly favorable setting for satellite data, and accuracy of the satellite predictions is currently higher than in many other countries. Second, the US already hosts one of the largest and possibly oldest area-based insurance scheme, based on county average yields. As we will show in more detail in the next section, lessons from index insurance schemes in the US are rather ambiguous. On one hand, for the main product from the Risk Management Agency (RMA), the demand for index insurance is very limited compared to the demand for farm-based insurance. On the other hand, the conclusion is reversed when one looks at another scheme offered by the Farm Service Agency (FSA). 


These opposite conclusions from different schemes motivate our main research question, seeking to measure the benefits of index insurance in a simplified world abstracting from institutional details, adverse selection and moral hazard. Results from our simulations suggest that index insurance performs surprisingly well compared to farm-based insurance. 
Having measured the benefits of index insurance, our second research question focuses on identifying the characteristics of the fields (or counties) that benefit most from index insurance. To do so, we extend the theory of \citet{Miranda1991} for the case comparing index insurance to farm insurance. 
We verify our new theoretical predictions on the dataset and find in general a solid agreement. 
A key result that arises from this exercise is that the determinants of the benefits of index insurance strongly depend on the choice of metric used. 
In particular, counties with high variance appear either to benefit the most from index insurance when compared to no insurance or the least when compared to farm insurance. To the best of our knowledge, this result is new in the literature, and we believe it has important implications for the design of index insurance. 

Our paper contributes to several strands of literature. For one, we contribute to the papers studying the US area-based insurance program \citep{Miranda1991,CarrikerWilliamsEtAl1991, SmithChouinardEtAl1994, BarnettBlackEtAl2005, DengBarnettEtAl2007} or similar area-based insurance programs elsewhere (\citealt{BreustedtBokushevaEtAl2008} for Kazakhstan and \citealt{YeHuEtAl2020} for China). Conclusions form this literature is rather mixed, with some papers suggesting that area-based insurance performs well compared to farm
\citep{BarnettBlackEtAl2005, DengBarnettEtAl2007} in the US, yet with counter-examples in China for example \citep{YeHuEtAl2020}. 
Compared to these papers, our sample size is orders of magnitude larger than any of these papers (the largest sample size we found is 66,000  in \citealt{DengBarnettEtAl2007}), but, more importantly, contains close to 600 insurance zones. This allows us to conduct a cross-county analysis of the determinants of the benefits of index insurance, something that, to the best of our knowledge, no other study has been able to do.
Our paper contributes also to the smaller literature estimating individual basis risk and its determinants. \citet{JensenBarrettEtAl2016} use a sample of 736 households in Kenya, and estimate basis risk and its determinants at the household level. They find that characteristics such as gender, the number of dependents in the household or the importance of livestock affect the basis risk. 
Compared to this study, we cannot observe household characteristics, but focus instead on field measures such as its mean and variance. 
Finally, we also make a contribution to the theoretical literature modeling the benefits of index insurance \citep{Miranda1991, Mahul1999, Vercammen2000, BourgeonChambers2003}. We extend the model of \citet{Miranda1991} to include a comparison of index insurance to farm insurance, instead of comparing index insurance merely to no insurance.

The paper is organized as follows: in Section~\ref{sec:Model}, we describe the Federal Crop Insurance Program, present our theoretical model and the utility metrics we use throughout. Section~\ref{sec:Data} presents the dataset, its construction and validation. Finally, we show our main results in Section~\ref{sec:Results}, and the robustness checks in Section~\ref{sec:Robustness}. 

\section{Context and conceptual model\label{sec:Model}}

\subsection{The US Federal Crop Insurance Program}

The US Federal crop insurance program has become since its inception in 1938 one of the largest programs of the Farm Bill, costing close to \$8 billions a year, second only to the nutrition program. These large costs can be explained by the generous nature of the program:  the government covers all operational costs, and subsidizes a large share of the premiums (40-60\%). These high subsidy rates are deemed necessary to induce farmers to participate into the program, given the relatively low  participation rates in early years. Participation is now high, with about 86\% of eligible acres covered in 2015. 

\begin{figure}

\begin{centering}
\includegraphics[width=0.95\columnwidth]{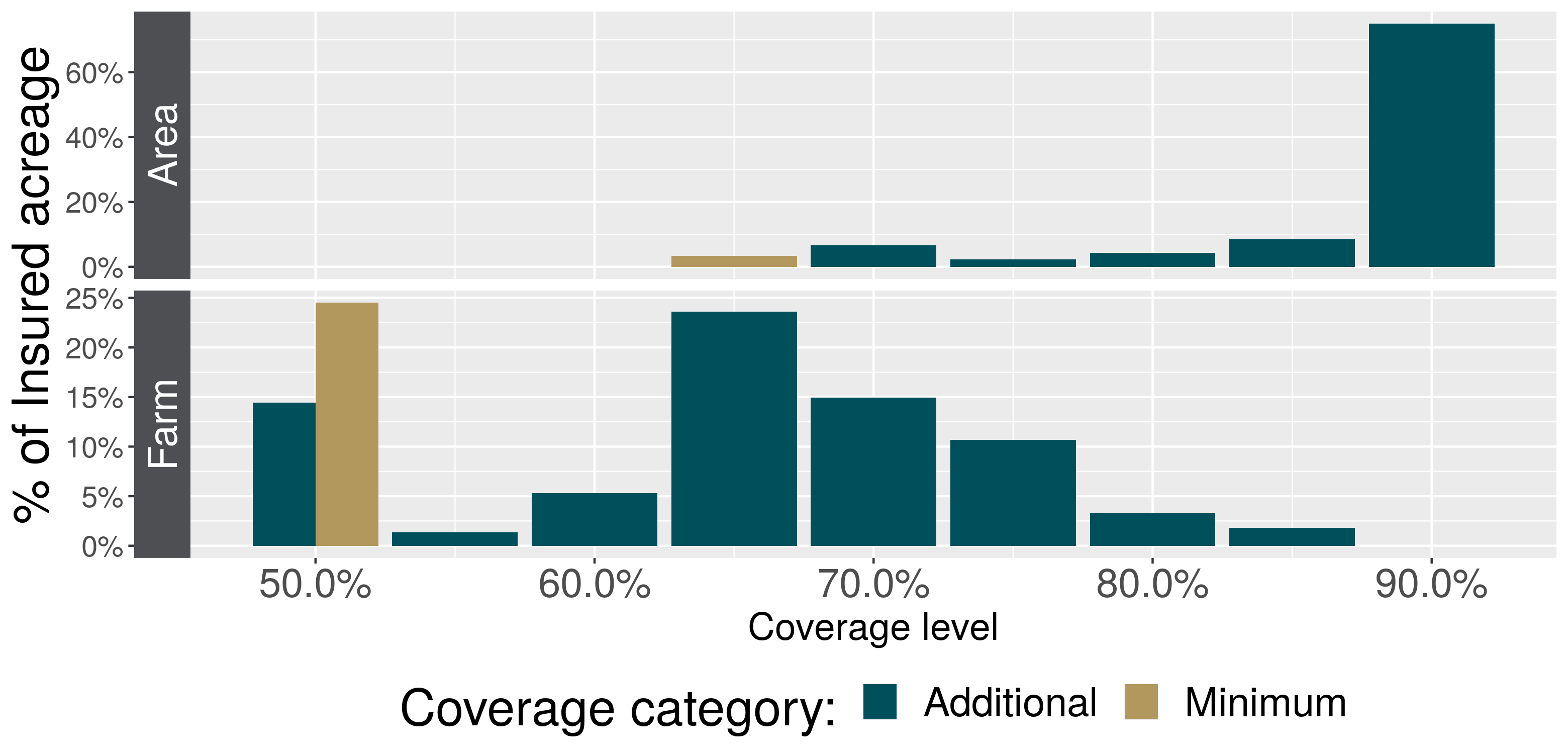}
\par\end{centering}
\caption{Demand for insurance at various trigger levels\label{fig:Demand-for-insurance}}

\tablecomment{Figure panels show the percent of total area under (top) area-based or (bottom) field-based insurance at different trigger levels. Source: Own computation from Risk Management Agency's Summary of Business}
\end{figure}

The Risk Management Agency (RMA) has offered a plethora of insurance plans throughout the years, with evolving names and specificities. 
In brief, these can be classified into plans insuring yields or revenue, and into plans insuring at the farm-level or at the county-level. Yield insurance at the farm level was historically the standard insurance plan. Area-based plans were introduced in 1993 under the initial name of Group Risk Plan. This area-based plan is an index scheme, where the index is the average county yield as measured by official statistics
collected by the US Department of Agriculture (USDA). The general idea behind all these plans is that indemnities are triggered whenever actual (farm or area) yield is below a certain percentage of its expected value. This \emph{trigger} level (called \emph{coverage} level in RMA terms) 
is offered at various levels, ranging from 50\% to 85\% for farm-level, and from 65\% to 90\% for the area-based product. Premiums are heavily subsidized, at an average rate of 60\%, with the rate decreasing for higher levels of trigger (see Table~\ref{tab:Subsidy-rate-for-area-farm} for details). Figure~\ref{fig:Demand-for-insurance} shows the trigger levels selected by the farmers for the farm- and area-based insurance over the 2011-2019 period averaging over corn and soybeans. The figure shows also the so-called catastrophic trigger (CAT) which comes at lower cost yet delivers lower indemnity. Strong differences appear between the farm and area-yield trigger selected. For the area-based scheme, the vast majority chooses the maximum trigger level, 90\%. On the other side, for the farm-based product, farmers choose either
the lowest trigger at 50\%, or an intermediate value of 65\%, while very few opt for the maximum coverage at 85\%.\footnote{The fact that farmers select only intermediate coverage for the farm-value has been discussed in various papers, see \citet{DuFengEtAl2017,Babcock2015,FengDuEtAl2020}} This difference between the trigger choice for farm- or area-based coverage suggests that area-based provides only a partial protection due to the basis risk. 

\begin{figure}

\begin{centering}
\includegraphics[width=0.95\columnwidth]{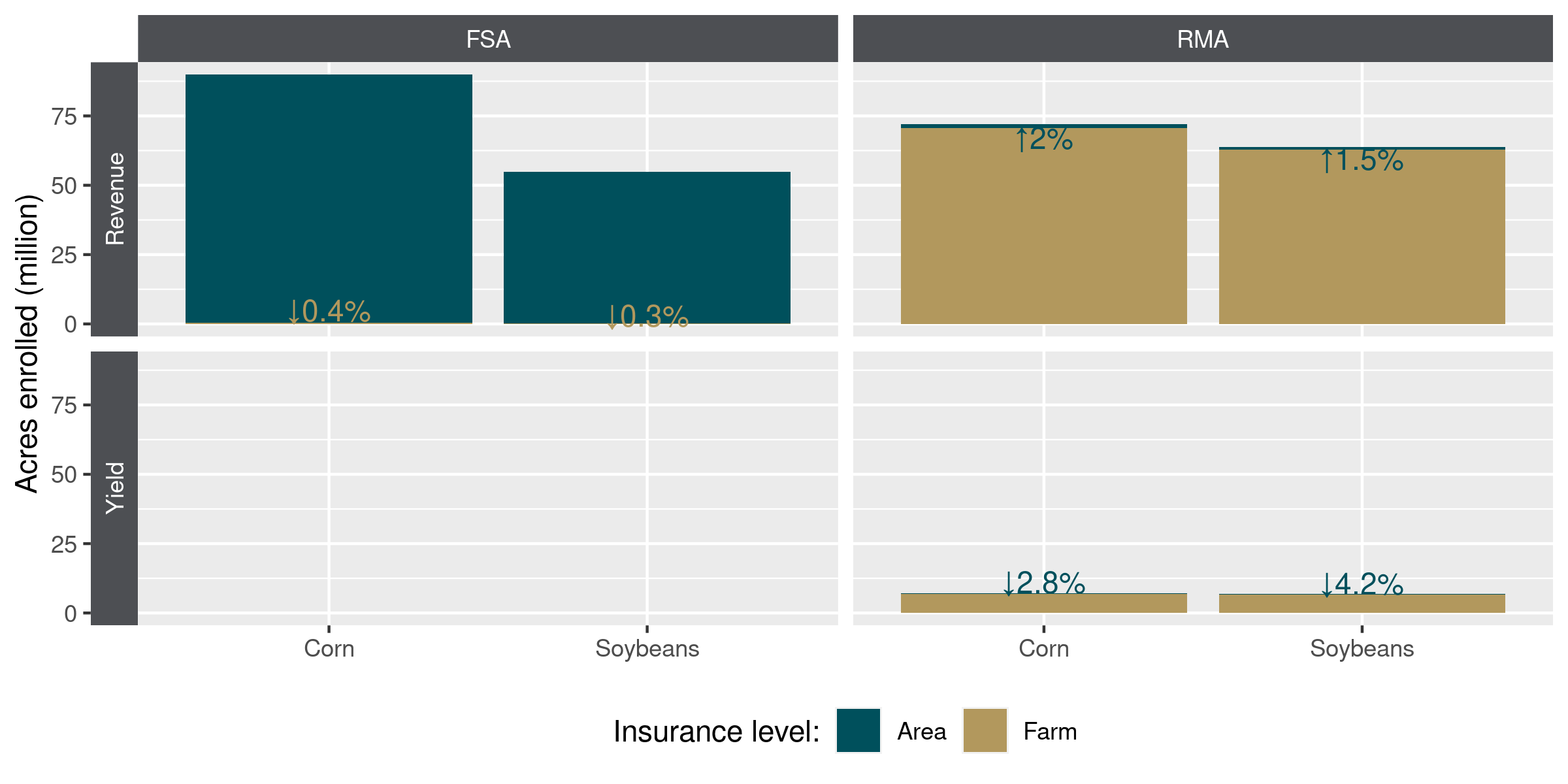}
\par\end{centering}
\caption{Demand for area versus farm-based insurance\label{fig:Demand-for-area-farm}}

\tablecomment{The panels show the acres enrolled in the area-based and farm-based programmes, both for the FSA (left) and RMA (right) products that cover either revenues (top) or yields (bottom). Source: Own computation from Risk Management Agency's Summary of Business}
\end{figure}

Do farmers prefer area- or farm-based insurance? The answer turns out to be rather ambiguous, depending on the program one considers.  Figure~\ref{fig:Demand-for-area-farm} shows the percentage of each scheme in terms of total acreage covered, both for the yield and revenue types.  The figure shows the demand both for the RMA product described above, as well as for a similar product offered by the Farm Service Agency (FSA), the Agriculture Risk Coverage (ARC) product. 
The demand for index insurance is very small for the RMA product, not more than 5\% in each case compared to traditional indemnity-based insurance. These results do not appear very encouraging for index insurance, casting doubt as to whether index insurance should be promoted at all. Interestingly, for the similar FSA product with both a farm- and area-based option, the conclusion is reversed: the area-based product is largely preferred over the farm-based one (see also \citealt{SchnitkeyCoppessEtAl2015}). 
There are of course plenty caveats in comparing directly these products, both across RMA and FSA, and within RMA products too. For the RMA comparison for example, not only are subsidy rates different for farm- or area-based schemes, but there are several other subtle differences that we sidestepped for the sake of clarity.\footnote{Most notably, we did not discuss here the details related to the \emph{protection price} for area-based insurance, nor the \emph{enterprise units} for farm-based insurance, all with different subsidy rates. Likewise, \emph{yield exclusion} options allowing to exclude a particularly bad year from the farm-level premiums increase the attractivity of farm-level products. } Furthermore, the comparison between FSA and RMA is difficult since the FSA product is a very recent product, and is offered at almost no cost.
Noting the difficulty to reach a firm conclusion on the benefits of index insurance from these apparently contradicting observations, we choose to run a large-scale simulation using a stylized representation of the RMA product abstracting from many institutional peculiarities. We choose in particular to rule out moral hazard and adverse selection, allowing us to focus on our main question of interest, the suitability and determinants of index insurance. 

\subsection{Conceptual model}\label{subsec:model}

Our objective here is twofold: we seek first to measure the benefits of index insurance, and then to understand the characteristics of fields (or counties) associated with those benefits. For each step, we compare the benefits of index insurance both to no insurance, as well as compared to farm-level insurance. We start here by presenting our model, and then discuss our choice of metrics for the benefit of index insurance.

Starting with some notation, we denote by $y_{ict}$ the yield for field $i$ in county $c$ at time $t$. We write the annual county average yield as $\bar{y}_{\cdot ct}$, the long-term county average yield as  $\bar{\bar{y}}_{\cdot c\cdot}$, where the $\cdot$ notation indicates over which dimension the averaging is done.\footnote{As an example, $\bar{y}_{\cdot ct}\equiv1/n_{i^c}\sum_{i\in c}y_{ict}$ is averaging over units $i$ in county $c$, and hence denotes the county mean over time.} For ease of exposition, we consider simplified payout schemes for both the area-based and farm-based insurances. The indemnity is triggered whenever the actual (farm or county) yield is below a long-term target, and the indemnity is simply the difference between the long-term target and the actual (farm or county) yield. For the area-based product, the long-term target is the county mean over time multiplied by a trigger level $\tau$, $\tau\bar{\bar{y}}_{\cdot c\cdot}$. The county indemnity $I_{ct}^c$ is simply the difference between the long term target $\tau\bar{\bar{y}}_{\cdot c\cdot}$ and the actual county yield, i.e. $I_{ct}^{c}=\max(\tau\bar{\bar{y}}_{\cdot c\cdot}-\bar{y}_{\cdot ct}, 0)$. This payout scheme differs slightly from the one used in practice,\footnote{The actual indemnity scheme divides the difference by the trigger $\tau_i$, and contains also a \emph{protection factor}, which allows to scale up or down the indemnity payment. The RMA does to our knowledge not provide data on insurance take-up by protection factor level, so we simply set it to 100\%, to ease comparison with farm-level insurance. See  \citet{SkeesBlackEtAl1997} for details.} yet this is the one used in \citet{Miranda1991} for which analytical results are available. Note that for ease of exposition the indemnity is expressed in yield units instead of dollars units. Scaling to dollar units is here unnecessary given that our focus is on yield, not revenue, insurance. Turning to the farm-based insurance, we consider  the same indemnity scheme, simply replacing county yields by individual field yields: $I_{ict}^{F}=\max(\tau_{i}\bar{y}_{ic\cdot}-y_{ict}, 0)$, where the long-term target is $\bar{y}_{ic\cdot}$, the field-level mean. 

\citet{Miranda1991} derived analytical expressions for the benefits of area-based index insurance versus no insurance. Further theoretical refinements of \citeauthor{Miranda1991}'s model focusing on the design of the optimal indemnity were made by \citet{Mahul1999}, \citet{Vercammen2000} and \citet{BourgeonChambers2003}. Miranda's model is based on $\beta_{ic}$, the coefficient of a regression over time of individual yields $y_{ict}$ against county yields $\bar{y}_{\cdot ct}$:

\begin{equation}
y_{ict}=\alpha_{ic}+\beta_{ic}\bar{y}_{\cdot ct}+\epsilon_{ict}\label{eq:MirReg}
\end{equation}

Intuitively, $\beta_{ic}$ indicates how well a farmer's yield is linked to the county yield. The term $\epsilon_{ict}$ represents idiosyncratic farmer-specific shocks that cannot be insured by a county-level insurance scheme. \citeauthor{Miranda1991} analyses the benefits of area-based insurance using a mean-variance utility framework. When premiums are fair, the difference in mean-variance utility between area-based and no insurance amounts to the difference in variance between the two schemes. \citeauthor{Miranda1991} shows that the variance reduction for field $i$ in county $c$,  $\Delta_{ic}^{\mycomp{C}{No}}$, is a function of the farmer's own $\beta_{ic}$, of the variance of the county indemnity $\sigma_{I^{C}}^{2}$ and of a county-level \emph{critical beta} value $\tilde{\beta}_{c}^C$: 

\begin{equation}
\label{eq:Mir_varRed}
\Delta_{ic}^{\mycomp{C}{No}}\equiv Var\{y\}-Var\{\tilde{y}^{C}\} =\sigma_{I^{C}}^{2}\left(\frac{\beta_{ic}}{\tilde{\beta}^{C}_{c}}-1\right)
\end{equation}

where $\tilde{y}^{C}$ denotes yields after county-based insurance, that is $\tilde{y}_{ict}^{C}=y_{ict}+I^C_{\cdot ct}-\pi^C$ (where $I^C_{\cdot ct}$ is the county-level indemnity and $\pi^C$ the premium).

It is easily seen from \eqref{eq:Mir_varRed} that there is risk reduction (equivalently, positive mean-variance utility from index insurance) for all fields whose $\beta_{ic}$ is above the critical beta, i.e. $\beta_{ic}>\tilde{\beta}_{c}$. Miranda shows furthermore that the critical beta is bounded between 0 and 1/2, and that the average of $\beta_{ic}$ is 1, suggesting that on average fields benefit from index insurance. Noting that the OLS coefficient $\beta_{ic}$ is related to the correlation coefficient $\rho_{ic}$ by $\beta_{ic}=\rho_{ic}\sigma_{ic}/\sigma_{c}$, we also see that within a county, the fields that will have a higher (mean-variance) utility form index insurance are those: 1) which are more correlated with the county average 2) which have a higher total variance $\sigma_{ic}$. The latter result we be particularly important later on, as it points towards a mechanical effect: a field's variance plays a large role in determining the benefits from index insurance, irrespective of its correlation to the index. The variance effect can outweigh the effect of the correlation with the index itself: a field that is mildly correlated with the index yet has a high variance can have a higher benefit from index insurance than a field that is perfectly correlated with the index yet has a small variance. 

Do the same insights hold when we compare index insurance to farm-based insurance? To investigate this, we derive first the expression for the variance reduction of a farm-level product versus no insurance, $\Delta_{ic}^{\mycomp{F}{No}}$. Yields with farm insurance, $\tilde{y}^{F}$, are given by  $\tilde{y}_{ict}^{F}=y_{ict}+I^F_{ict}-\pi^F$, where $I^F_{ict}$ and $\pi^F$ are respectively the indemnity and premium. Note that at this point, we have kept the trigger level $\tau$ implicit. The variance of the yields is given by: $\Var\{\tilde{y}^{F}\}=\sigma^2_{ic}+\sigma^{2}_{I_{ic}^F}+2\Cov(y_{ict}, I_{ict}^F)$. Following the same reasoning as in \citet{Miranda1991}, we define the \emph{farm-level critical beta} $\tilde{\beta}^F_{ic}=-\Var\{I^{F}_{ict}\}/2\Cov(y_{ict},I^{F}_{ict})$. We can now rewrite the variance of yields with farm insurance as:

    $$Var\{\tilde{y}^{F}\}=\sigma^2_{ic}+\sigma^{2}_{I_{ic}^F}\left(1-\frac{1}{\tilde{\beta}^{F}_{ic}}\right)$$
    
The variance reduction from a farm-level product $\Delta_{ic}^{\mycomp{F}{No}}$ compared to no insurance is then: 

\begin{equation}
\label{eq:Delta_farm_vs_none}
\Delta_{ic}^{\mycomp{F}{No}}\equiv Var\{y\}-Var\{\tilde{y}^{F}\}=\sigma^{2}_{I_{ic}^F}\left(\frac{1}{\tilde{\beta}^{F}_{ic}}-1\right)
\end{equation}

This expression is very similar to the variance-reduction from index insurance at the county level $\Delta_{ic}^{\mycomp{C}{No}}$ in (\ref{eq:Mir_varRed}), the main difference being that 
 the $\beta_{ic}$ coefficient vanished (it is now 1). The variables $\sigma^{2}_{I_{ic}^F}$ and $\tilde{\beta}^F_{ic}$ keep the same interpretation as in the Miranda model, yet refer to the individual yield instead of the county average yield. In particular, the bound on the critical beta $\tilde{\beta}_{ic}$ between 0 and 1/2 also holds, implying that the benefit from farm-level insurance is $\geq0$.

Determining formally the effect of a field variance on the utility of farm insurance is not trivial. The expression above involves mixtures of truncated variables, for which analytical expressions become quickly cumbersome. We computed the formula for the normal distribution, and verified numerically the effect of increasing the field's mean and variance, as well as the trigger parameter $\tau$. Without surprise, our computations show that increasing a field's variance or the trigger level $\tau$ both strengthen the variance reduction $\Delta_{ic}^{\mycomp{F}{No}}$. On the other hand, the impact of the field mean $\mu$ depends on the trigger level $\tau$: for $\tau>1$, increasing $\mu$ accentuates the variance reduction, while for $\tau<1$ is reduces it. Finally, for $\tau=1$, changing $\mu$ has no impact on the variance reduction. 

Having obtained this expression, we can finally derive the variance reduction from area-based insurance compared to farm-insurance:

\begin{equation}
\label{eq:Delta_area_vs_farm}
\Delta^{\mycomp{C}{F}}_{ic}\equiv
Var\{\tilde{y}^{F}\}-Var\{\tilde{y}^{C}\}=\sigma^{2}_{I_{ic}^F}\left(1-\frac{1}{\tilde{\beta}^{F}_{ic}}\right)+\sigma^{2}_{I_{c}^C}\left(\frac{\beta}{\tilde{\beta}^{C}_{c}}-1\right)
\end{equation}

Noting that the first term  in \eqref{eq:Delta_area_vs_farm} corresponds to $-\Delta_{ic}^{\mycomp{F}{No}}$ in \eqref{eq:Delta_farm_vs_none} and the second term to $\Delta_{ic}^{\mycomp{C}{No}}$ in \eqref{eq:Mir_varRed}, we  can rewrite the variance reduction of area-based insurance versus farm as: $\Delta^{\mycomp{C}{F}}_{ic}=-\Delta^{\mycomp{F}{No}}_{ic}+\Delta^{\mycomp{C}{No}}_{ic}$. This expression is particularly useful to sign the effect of the farm-level trigger $\tau$, the field mean $\mu_{ic}$ and the field variance $\sigma^2_{ic}$. Without surprise, increasing the trigger level $\tau$ for the farm insurance dampens the variance reduction of farm-based insurance versus farm insurance $\Delta^{\mycomp{C}{F}}_{ic}$. This is to be expected: when we compare index insurance to increasingly high coverage of farm-based insurance, the benefits of index insurance diminish. The impact of the field mean $\mu_{ic}$ depends on the magnitude of $\tau$. Assuming that $\tau<1$, as is the case in practice, $\mu_{ic}$ is expected to increase $\Delta_{ic}^{\mycomp{C}{F}}$, given that it reduces $-\Delta_{ic}^{\mycomp{F}{No}}$. The impact of the field variance $\sigma^2_{ic}$ is theoretically ambiguous: on one hand, $\sigma^2_{ic}$ reduces the first term (given that it increases $\Delta_{ic}^{\mycomp{F}{No}}$), but on the other hand it increases the second term, given that $\beta_{ic}=\rho_{ic}\sigma_{ic}/\sigma_{c}$. Lastly, the correlation coefficient $\rho_{ic}$ has a positive sign as one would expect, given that increasing this coefficient increases the benefit of index insurance in general.



All the analytical results so far focused on variance reduction, which is equivalent to an increase in mean-variance utility. The use of a mean-variance utility function is, however, somehow controversial. \citet{JensenBarrettEtAl2016} argue in particular that the assumption of symmetry in preference between positive and negative shocks is not very relevant for the context of crop insurance, targeted at reducing negative shocks. Expected utility (EU) offers a theoretically-grounded alternative, and captures the asymmetry in preferences through the concavity of the risk aversion function. For this reason, we use EU-based measures in this paper, as detailed in the next section. 

How can we transpose our analytical results based on a mean-variance utility to a utility of general form? This can be done by using a second-order Taylor expansion of the expected utility function. As detailed in Proposition \eqref{th:taylor} in the Appendix, a second-order Taylor approximation of the difference in expected utility between scheme $A$ and $B$ gives: $E[u(A)]-E[u(B)]=1/2 u''(\mu)\left(\sigma^2_{y^A}-\sigma^2_{y^B}\right)=-1/2 u''(\mu)\Delta^{\mycomp{A}{B}}$.
This means that the difference in expected utility between two schemes is proportional to the variance reduction computed above. In particular, the benefit of index insurance $C$ versus none is $E[u(C)]-E[u(No)]=-1/2 u''(\mu)\Delta^{\mycomp{C}{No}}$, where $\Delta^{\mycomp{C}{No}}$ is the variance reduction factor derived by \citet{Miranda1991}. Likewise, the benefit of index insurance versus farm insurance $F$ is $E[u(C)]-E[u(F)]=-1/2 u''(\mu)\Delta^{\mycomp{C}{F}}$, where $\Delta^{\mycomp{C}{F}}$ is the variance reduction derived in equation \eqref{eq:Delta_area_vs_farm}. This result implies that except for the parameter $\mu$, the signs of $\sigma^2$, $\tau$ and $\rho$ will remain the same as per the discussion based on mean-variance utility (as $u''$ is negative following standard expected utility assumptions). For the parameter $\mu$ on the other hand, note that now we need to take into account the additional term $u''(\mu)$. Assuming the utility function exhibits prudence ($u'''>0$), the effect of $\mu$ on the first term becomes negative and hence could potentially reverse the effect of $\mu$ discussed above.

The reader might notice that we haven't mentioned the traditional concept of basis risk in our theoretical results. This seems to be at odds with the oft-repeated idea that basis risk, i.e. the probability that a farmer experiences a loss, while the index does not leads to a payout, is the main issue with index insurance.\footnote{The reverse situation of the farmer experiencing no loss yet receiving an indemnity is also possible, but usually not taken into account, as the emphasis is on the ability of an insurance scheme to reduce negative events, not to amplify positive ones. It should be noted however that such insurance windfalls have also an indirect negative impact by increasing premiums.} This ''loss without indemnity'' definition of basis risk, also called False Negative Probability by \citet{ElabedBellemareEtAl2013}, is written as: 

\[
FNP(\theta_{c},\theta_{i})\equiv P(\bar{y}_{\cdot ct}>\theta_{c}|y_{ict}<\theta_{i})=\frac{P(\bar{y}_{\cdot ct}>\theta_{c}\cap y_{ict}<\theta_{i})}{P(y_{ict}<\theta_{i})}
\]


Here $\theta_{c}$ is a county loss threshold, 
and $\theta_{i}$ is a farmer-specific subjective loss threshold. This measure is unfortunately unsatisfying for multiple reasons. First of all, it requires to define  specific loss thresholds $\theta_{c}$ and $\theta_{i}$, which is mostly arbitrary given that yields are a continuous variable. Second, this is only a probability between 0 and 1, and hence is not indicative of the amount of loss experienced. An insurance missing a particularly catastrophic event yet delivering payouts for all other small loss events would be deemed to offer a low basis risk despite not serving when it is the most needed.\footnote{See \citet[see ][]{Clarke2016,BarreStoefflerEtAl2016} for an in-depth discussion of metrics for index insurance.} Furthermore, estimation of this quantity can prove unstable unless ones makes distributional assumptions, as one is essentially doing a division involving the probability of a rare event in the denominator. The lower the threshold used for the individual loss, the more unstable this quantity can become, a phenomenon we observe in this sample (see Table~\ref{tab:basis-risk-by-level}).

Given the limitations of the ''loss without indemnity'' definition of basis risk, we follow the insights from our theory developed above and focus on an alternate definition of basis risk related to the $R^2$ of the field-to-county regression \eqref{eq:MirReg}. Following, \citet{ElabedCarter2015}, we look at the variance of residuals $\sigma_{\epsilon_{ic}}^{2}$ normalized by the field-specific variance, which is equivalent to $\sigma_{\epsilon_{ic}}^{2}/\sigma_{ic}^{2}=1-R_{ic}^{2}$. This represents the amount of idiosyncratic risk that cannot be insured by the index. Like the false negative probability, this measure of basis risk is between 0 and 1. A value of 0 indicates perfect correlation with the index, while a value of 1 indicates that the variables are fully uncorrelated. While this is our preferred measure of basis risk, we will still include the ''loss without indemnity'' measure in the regressions below for the sake of comparison. 

\subsection{Empirical measures of the utility of index insurance and of basis risk }\label{subsec:measures}

Our objective here is to measure the benefits of index insurance. We seek to compare index insurance both to no insurance and to farm-based insurance. The comparison to farm-based insurance is a more stringent comparison, keeping in mind that ``no insurance'' can be thought as a farm-based insurance with a trigger of 0\%. When comparing to farm insurance instead, we are raising this trigger (to values above 50\%), increasing the strength of our comparison. 

To compare the different plans, we use expected utility (EU) measures, assuming a functional form for the utility function, and evaluating the function over yields without insurance, with index insurance and with farm-based insurance. Following previous literature \citep{WangHansonEtAl1998,DengBarnettEtAl2008,FlatnesCarterEtAl2018} we use a constant relative risk aversion (CRRA) iso-elastic utility function, with a parameter of 1.5. Fair premiums and indemnities are computed ex-post from the data. By following this procedure, we make two fundamental assumptions. Firstly, we are assuming that yields are the same whether or not the farmer takes insurance. This means that we are ruling out possible moral hazard. Second, we are computing ex-post fair premiums assuming the farmer takes the insurance every period, ruling out adverse selection. While this makes us depart from real-world characteristics in an important way, this allows us to focus on our main topic of interest, the utility of index insurance. 

Having obtained measures of expected utility for each scheme, we need to compare them across schemes. We do this in two different ways. For the comparison of index insurance versus no insurance, we do our comparison in yield units using certainty equivalents (CE) measures instead of directly comparing EU values. The certainty equivalent is the non-random value whose utility is the same as the expected utility from a random \emph{lottery}, where the lottery here is simply the set of observed yields. That is, CE is the value such that $u(CE) = E[U(y)]$ holds.  A higher CE is equivalent to a higher utility, and hence we simply compare index insurance versus no insurance based on their ratio $CE^{C}/CE^{no}$. A ratio >1 implies a higher utility of index insurance, $U^{C}>U^{no}$. We will also use the concept of \emph{risk premium}, which is simply the difference between the mean of a distribution and the certainty equivalent, $RP\equiv \mu-CE$. When comparing two schemes, a lower risk premium is equivalent to a higher utility.


For the comparison of index insurance versus farm insurance, we are faced with the additional problem of having to choose the trigger level for the farm insurance. As discussed above (see Figure~\ref{fig:Demand-for-insurance}), while almost all farmers choose the 90\% trigger level for area-based insurance, there is much more heterogeneity for the choice of the farm-level trigger. To handle this, we introduce a new measure describing the equivalent in terms of farm coverage of an index-based product with coverage at 90\%. We name such measure the \emph{farm-equivalent risk coverage}, which we define as the highest level of farm-based insurance for which index insurance is at least as good or better. The higher this number is, the more protection index insurance gives in terms of an ideal farm-based scheme. Formally, our measure is defined as:

\[
\tau^{*}\equiv\max_{\tau\in\{0.2,\ldots,0.9\}}\tau\,\quad\textrm{such that}\ensuremath{\quad}U_{90\%}^{area}>U_{\tau\%}^{farm}
\]

We set the value of 90\% for the area insurance as this is the value most selected by farmers, and search over a large set candidates values of $\{0.2,\ldots,0.85,0.95\}$ which includes all values offered by the RMA (from 0.5 to 0.85). For an index-insurance with a trigger of 90\%, this measure will typically lie in the interval $[0\%, 85\%]$. We expect indeed that at an equal trigger level, a farm-based insurance at 90\% will be preferred to an area-based insurance at 90\%, given that the farm plan will additionally cover the idiosyncratic risk on top of the systematic covered by the index insurance. But this is not necessarily the case, and we observe several cases where the area insurance at 90\% is preferred to a farm-based product at 90\%, or even at 95\%. To understand this counter-intuitive situation, think of a field perfectly correlated to the county area except for the only period when there is a large shock: say the field experiences a shock of say 84\% compared to its mean, while the county average has a shock of 70\% (we will assume there are enough periods so that the mean of the county is close to the field mean despite the one-time discrepancy). A farm-based product with coverage of 85\%, 90\% or 95\% will provide an indemnity of 1\%, 6\% or 11\% respectively. On the other side, the indemnity of the area product will be close to 20\%. In this case the farmer will clearly prefer the area-based product at 90\% to a farm product at 90\%, and might in fact even prefer it compared to a farm product at 95\%!  Figure~\ref{fig:yields_pre_post_1plot} in the appendix shows an example of a field in our sample which has a farm-equivalent risk coverage of 90\%. This is explained by a very high indemnity from the area product, at 110\% of own yields, for the lowest yield realization (at 70\%).

A limitation of our \emph{farm-equivalent risk coverage} measure is that it is undefined in two cases. For one, if index insurance is not even as good as no insurance, then it is clear that it won't be better than any level of farm-based insurance. In this case, we attribute a $\tau$ value of 0\%. The second limitation arises from the fact that we can only observe the utility of farm-level insurance for those trigger levels at which there is a yield fallout happening. If for a given field the minimum yield observed is say at 70\% of the average yield, then farm plans covering 50\% to 65\% will not provide any protection, and hence will give the same utility as the situation without insurance. If however index insurance is inferior to the minimum observed relative yield, we only know that it lies in an interval $[0, y_{\min}/\bar{y}[$. These two limitations raise issues for the aggregation of our \emph{farm-equivalent risk coverage} at the county level. To address these two issues, we consider rank statistics such as the median and the proportion of fields within a county which have $\tau_i>0.85$, as well as $\tau_i>0.50$.\footnote{Using rank statistics will take care of the issue of aggregating over zero values. It will also partially address the problem of undefined values, although there is still a small percentage of fields with relative minimum above 85\% (or above 55\%) which would be miscounted. The total percentage of undefined values is however relatively small, between 2\% and 5\%.} These numbers correspond to the highest and lowest levels of farm protection available. The 50\% level is also called \emph{catastrophic} protection, so serves as  a good benchmark for the minimum protection level index insurance should provide. The 85\% level on the other side corresponds to the best available farm-level protection, so any field with a farm-equivalent risk coverage at 85\% or above would strictly prefer area-based insurance over far-based insurance.


\section{Data\label{sec:Data}}

The yield data comes from the Scalable Satellite-based Yield Mapper (SCYM) model initially developed by \citet{LobellEtAl2015} and further refined in \citet{JinAzzariEtAl2017}.  The SCYM method predict yields based on a satellite-derived vegetation index.\footnote{The model uses the so-called Green Chlorophyll Vegetation index (GCVI) which is similar in spirit to the widely known normalized difference index, NDVI.} The model follows an innovative approach using an agronomic crop growth model to create a training set. In brief, the agronomic model is used to simulate multiple realizations of \emph{pseudo} yields and \emph{pseudo} vegetation values. These simulated pseudo values are used to train a regression between vegetation index and yields. The estimated parameters are then used to predict yield based this time on the satellite-observed vegetation index. The advantage of this method is that it does not require ground data for the training stage. When ground truth data is available, it can be used as true out-of-sample validation. When validated against ground truth data for more than twenty thousand corn fields, \citet{DeinesDadoEtAl2019} find that the overall correlation for corn is 0.67 at the field level. Accuracy for soybeans is typically lower, between 0.5 and 0.6. Earlier versions of this dataset have been used in various studies, \citet{LobellAzzari2017} look at increasing field heterogeneity over time, \citet{SeifertAzzariEtAl2018} study the effect of cover crops, \citet{DeinesWangEtAl2019} study the effect of conservation tillage, \citet{Stigler_rotation_2019} estimates the effect of rotation and \citet{Stigler2018} measures the supply response to prices.

We extend here this dataset in several ways. Firstly, we use an improved version of the model based on recent developments by \citet{DeinesDadoEtAl2019} for corn and \citet{DadoDeinesEtAl2019} for soybeans. Second, we extend the sample over time. Previous versions of the dataset were predicting yields for those pixels designated as corn and soybeans by the Cropland Data Layer (CDL) from \citet{BoryanEtAl2011}. The CDL crop map covers our nine states of interest in the Corn Belt, yet starts at different periods depending on the state, with starting dates ranging from 2000 to 2008. To have a consistent sample, we fill-in the missing years with the Corn-Soy Data Layer (CSDL) crop map of \citet{WangDiTommasoEtAl2020}. The CSDL crop map uses random forests to predict corn and soybean from 2000 onwards for the nine states we consider here. To measure the accuracy of our resulting dataset, we aggregate the SCYM yield predictions at the county level, and compare those against NASS county means. We find a correlation between the our predicted values and official statistics of 0.81 for corn and 0.77 for soybeans. 

The SCYM yield dataset is at the pixel level, which is not very relevant to our analysis here. We use a dataset of field boundaries, the Common Land Unit (CLU). This dataset was last available for the year 2008, so field shapes might have changed in the meanwhile. To address this, we select only fields for which there is a good classification agreement throughout the years. That is, we compute the frequency of the majority vote from the crop pixel classification map (CDL or CSDL), and retain only fields that have on average 85\% of classification agreement. Put differently, we require that for a given field, the CDL consistently predicts either corn or soybean for 85\% of the pixels throughout the years. The pixel count is done by only considering pixels within a negative 30m buffer, avoiding contamination by mixed pixels lying on the border of the field. Averaging of the SCYM yields within the field boundary is made using the same subset of interior pixels. 

In addition to filtering field boundaries based on classification agreement, we also restrict the sample to consider field-crop pairs that have at least eight years of observations of corn or soybeans. This is to guarantee statistical accuracy when we estimate the field-to-county regressions. This implies that we might observe a field for only one crop or for both. Applying these two filters, we are left with 1,838,199 million fields in the 9 states we consider. Among these 1.8 millions fields, for 54\% of those we observe data on both crops (i.e. we have more than $\geq8$ observations for each of corn and soybeans), while for 28\% we observe only corn and 18\% only for soy. The resulting dataset gives us 2,826,681 million field-crop pairs. The sample runs from 2000 to 2019, and the total size of our sample is $\sim 30$ million field-crop-year observations. 


A defining characteristic of production in the Corn Belt is the practice of rotation between corn and soybeans \citep[see ][]{Hennessy2006,SeifertEtAl2017}. Given the large dataset we have, we observe almost every possible sequences of corn and soybean (and other crops), from always corn, always soy, always rotating to any other intermediate combinations. This raises a problem for the computation of fair premiums for the area-based product. Our fair premiums are computed using county yields for the whole period. This means that the premiums will be fair for fields planting always corn (or always soy) over the whole period. But for other fields, the premium might be exceptionally favorable (say field is planted to corn only in drought year 2012 and receive huge indemnity) or very unfavorable (field is planted to corn every year but 2012). This brings important randomness in our data, making our comparisons blurred. To avoid this, we resort to simulating yields, which provides us with a sample of corn and soy yields every year for each field. This has three further advantages.  First of all, this allows us to extend the time length of our sample, which we simulate using NASS means from 1990 to 2018. Second, having more observations for each field increases the probability of observing lower  minimum values for each field, attenuating the problem of undefinedness of our farm-equivalent measure, which is not defined if the observed relative minimum is too high. Finally, simulating data can be seen as a measurement error correction, where we adjust our sample to match official county means.  Yields are simulated based on the field-to-county regression~(\ref{eq:MirReg}): for each field we estimate $\hat{\alpha}_{ic}$, $\hat{\beta}_{ic}$ and $\hat{\sigma}_{ic}^{2}$ based on the raw data. We then plug-in detrended NASS county means $\tilde{y}_{ct}^{NASS}$ from 1990 to 2018, and simulate residuals from a normal distribution $\mathcal{N}(0,\hat{\sigma}_{ic}^{2})$, that is $\hat{y}_{ict}\sim\mathcal{N}(\hat{\alpha}_{ic}+\hat{\beta}_{ic}\tilde{y}_{ct}^{NASS},\hat{\sigma}_{ic}^{2})$. To avoid simulating outlying observations, we actually simulate using a truncated normal distribution, setting generous lower bounds of 10 [bu/acres] for both crops, and upper bounds of 100 [bu/acres] for soy and 350 [bu/acres] for corn. As a robustness check, we investigate the impact of using raw data instead in Section~\ref{subsec:robust_simulated}. 

\section{Results\label{sec:Results}}

\subsection{Descriptive statistics}

\begin{figure}

\subfloat[Coefficient of variation\label{fig:Yield's-coefficient-of-variation}]{

\begin{centering}
\includegraphics[width=0.95\columnwidth]{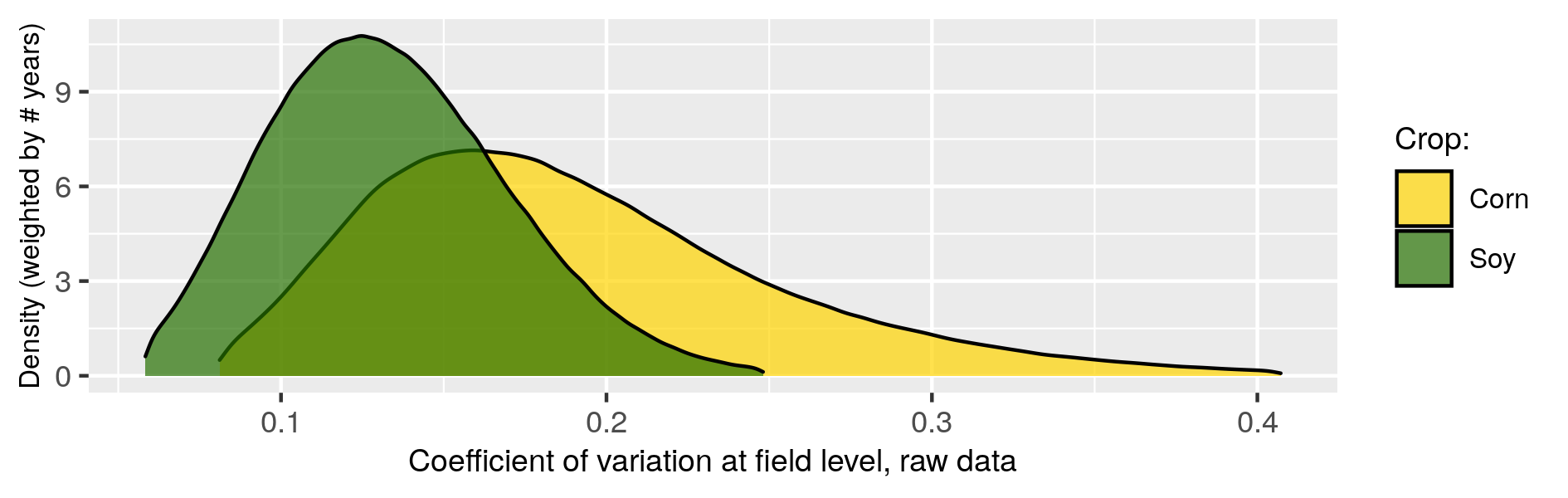}
\par\end{centering}
}

\subfloat[Basis risk $1-R^{2}$\label{fig:basis_risk_density}]{

\begin{centering}
\includegraphics[width=0.95\columnwidth]{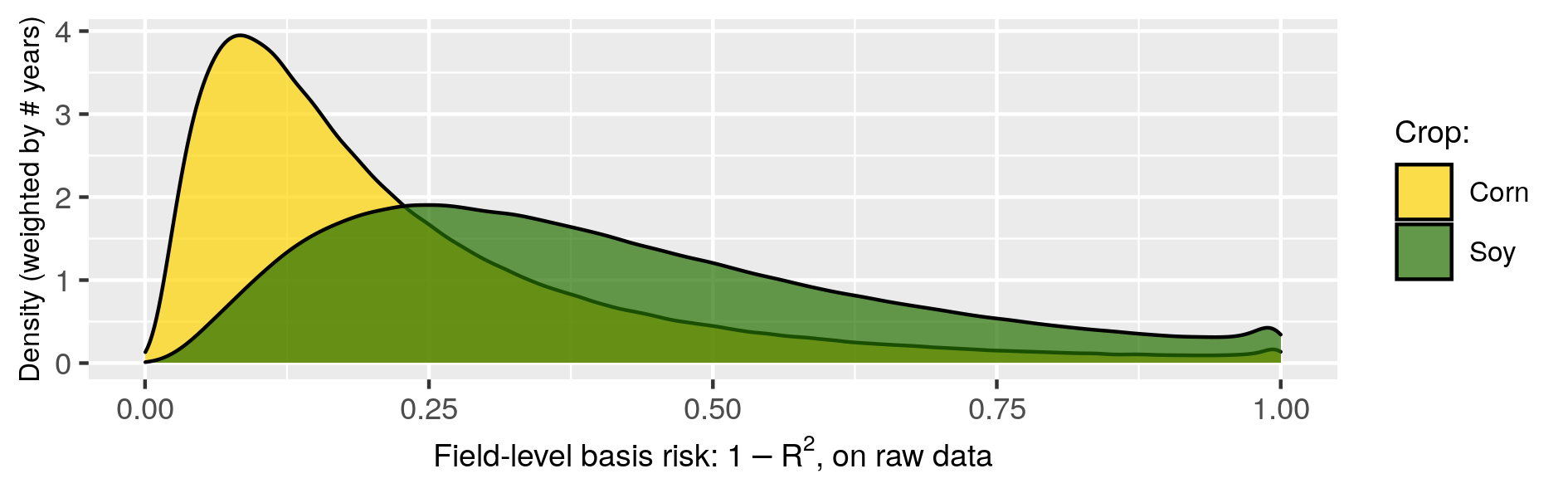}
\par\end{centering}
}

\caption{Field level total and idiosyncratic variability (raw data)}

\tablecomment{Source: own computation from SCYM. We kept only fields that have at least 8 observations for each crop. The density is further computed by weighting fields by the number of years planted to the respective crop.}
\end{figure}

We start here by discussing some figures and descriptive statistics of the temporal variance and the basis risk based on the raw (un-simulated) sample. Figure~\ref{fig:Yield's-coefficient-of-variation} shows the field-level total variance over time  and \ref{fig:basis_risk_density} shows its idiosyncratic part. The temporal variance is expressed using the coefficient of variation for ease of comparison across crops. The basis risk measure is the $1-R^{2}$ from the field-to-county regression \eqref{eq:MirReg}. While corn appears to be more variable than soybean (panel a), its variability is better captured by co-movement with county averages, so that the idiosyncratic variability is much lower than for soybeans (panel b).

Table~\ref{tab:sample-desk-stats} shows descriptive statistics for the same sample. We have slightly more fields with at least 8 years of corn than we have for soybeans. Corn is also  planted slightly more often than soy, which is mainly due to the fact that almost no fields are cultivated exclusively with soy, while a larger share follows corn mono-cropping \citep{Stigler_stlyised_CB}. The average of the coefficient of variation is higher for corn than for soybeans, confirming the visual impression from Figure~\ref{fig:Yield's-coefficient-of-variation}.  Consistent with this observation, we observe that the \emph{average relative minimum yield}\footnote{The \emph{average minimum relative yield} is averaging over each field's minimum divided by the respective field's mean.} is lower for corn, and that the average probability of having yields lower than 75\% is higher for corn. Turning to measures of basis risk, we show the ''loss without indemnity'' measure as the probability of experiencing a a loss below 75\% but no indemnity from a 90\% county coverage, and our preferred measure based on the $1-R^2$ from regression \eqref{eq:MirReg}. According to both measures, soybeans has a much higher basis risk compared to corn. As the threshold of 75\% to assess loss is arbitrary, Table~\ref{tab:basis-risk-by-level} in the appendix shows the basis risk varying the loss thresholds. At 85\% --the highest coverage offered for farm-based insurance-- the risk of not receiving an indemnity is relatively low, at  20\% for corn, yet as high as 36\% for soybeans. The risk decreases for lower loss thresholds, and become negligible (3\%) for corn when assessed against losses below 65\%. For soybeans, this number does not monotonically decrease and instead increases at very low thresholds. This is likely due to the fact that there are very few fields that experience losses below 65\% (less than 0.5\%), rendering our estimates unstable, as indicated by the wider confidence intervals.

\begin{table}[ht]
\centering
\caption{Sample descriptive statistics} 
\label{tab:sample-desk-stats}
\begin{tabular}{llll}
  \toprule
Category & Variable & Corn & Soy \\ 
  \midrule
Counts & Number of fields with at least 8 years & 1 498 803 & 1 327 878 \\ 
   & Average number of years planted & 10.7 & 10.0 \\ 
  Yields & Mean yield [bu/ac] & 162.1 & 51.6 \\ 
   & Average CV (over time) yield  & 0.19 & 0.13 \\ 
   & Average relative minimum yield & 63.3\% & 78.7\% \\ 
   & P(field $<$ 75\%) & 7.5\% & 3.1\% \\ 
  Basis risk & P(county$>$90\% $|$ field$<$75\%) & 6.8\% & 24.7\% \\ 
   & Basis risk (1-$R^2$) & 0.23 & 0.41 \\ 
   \bottomrule
\multicolumn{4}{l}{Source: raw dataset. Last three rwos are averages over fields.}\\
\end{tabular}
\end{table}

\subsection{Utility of index insurance at the field level}

We turn now to the measures of the utility of area-based insurance. These measures are based on the simulated dataset, which contains 29 years of data for each field. We always use the 90\% coverage level for the area-based scheme, since this is the level overwhelmingly chosen in practice (see Figure~\ref{fig:Demand-for-insurance}). We first measure the raw cost of risk using the concept of \emph{risk premium}, which is simply the difference between the mean of the yields and the certainty equivalent. For a given aversion parameter, the higher this difference, the higher the cost of risk. We find that risk induces a welfare loss in yield metric of 7 [bu/acre] for corn and 0.6 [bu/acre] for soy, amounting to 4.3\% and 1.2\% in percentage compared to the mean. The higher cost of risk for corn is consistent with the higher variability noted above.

\begin{figure}

\centering{}\includegraphics[width=1\columnwidth]{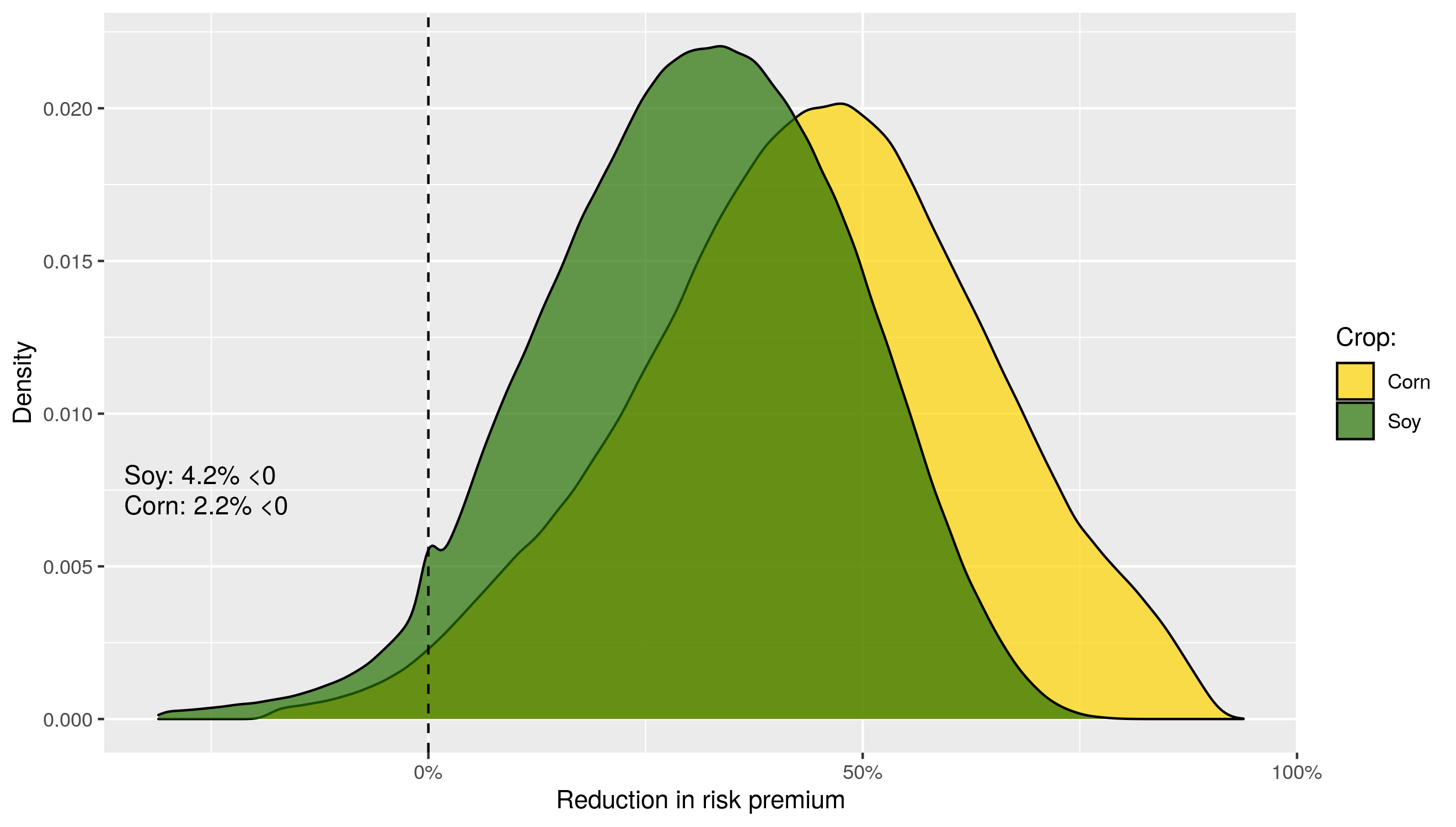}

\caption{Comparing index insurance versus none: reduction in risk premium \label{fig:Comparing-CE_vs_none}}

\tablecomment{Source: simulated data. }

\end{figure}

How effective is index insurance at reducing the risk premium? Figure~\ref{fig:Comparing-CE_vs_none} shows the density of the risk premium reduction, which indicates how the risk premium with no insurance is reduced with index insurance. A higher reduction means a higher utility from index insurance. We see clearly that index insurance provides more risk reduction for corn compared to soybeans. The average reduction for corn is 43\%, while that for soybeans is about 30\%. 
The proportion of fields for which the reduction is negative, i.e. whose utility from index insurance is actually lower than without insurance, is also lower for corn, at 2.2\% against 4.2\% for soybeans. The better results for corn are in line with the findings from Figure~\ref{fig:basis_risk_density} and Table~\ref{tab:sample-desk-stats} above, which showed that the basis risk was lower for corn. The small clustering of values around 0 for soybeans arises from the presence of a county where the county average was never much below 90\%. In this case, indemnities were so small that the utility and risk premiums with or without insurance are very close.

\begin{figure}

\centering{}\includegraphics[width=1\columnwidth]{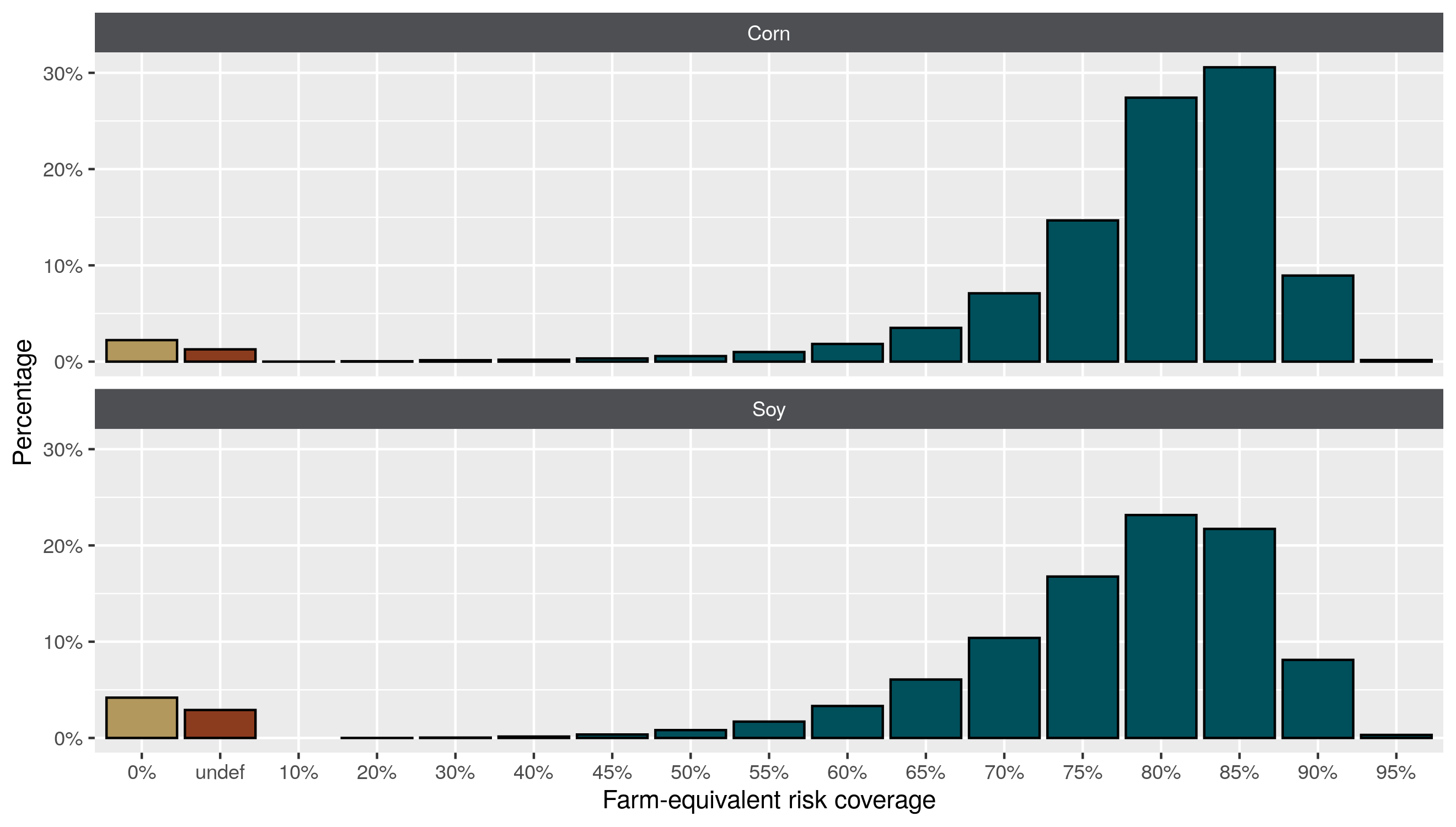}
\caption{Comparing index insurance versus farm insurance \label{fig:Comparing-CE_vs_farm}}

\tablecomment{Source: simulated data.}
\end{figure}

Moving to the more stringent comparison between index insurance and farm insurance, Figure~\ref{fig:Comparing-CE_vs_farm} shows our measure of \emph{farm-equivalent risk coverage}. The category $0\%$ indicates fields for which index insurance is not even as good as no insurance. These are the same percentages as the  ones in Figure~\ref{fig:Comparing-CE_vs_none}. The category \emph{undef} corresponds to those fields for which our comparison is undefined given that the utility of farm-based insurance can only be observed for those coverage levels above the observed minimum yield (see discussion in \ref{subsec:model}). This amounts to a small percentage of observations, a fact that is due to our use of the simulated data which tends to contain lower minima than the raw dataset.\footnote{To see the impact of using the raw dataset instead, see the robustness check in \ref{subsec:robust_simulated}.} Focusing on the subset of well-defined values, it is apparent again that corn provides a higher protection than soybeans. Looking at our measure of fields with at least an equivalent coverage of 85\%, this number is relatively high, at 40\% and 30\% percent respectively. 
This is an important result, as it suggests that index insurance performs quite well relatively to the best available farm-level insurance level. Another interesting level to compare to is the 50\% trigger, which is also the so-called \emph{catastrophic} level offered at very low cost for farm-level insurance. There is now a large amount of fields for which index insurance is at least as good as this 50\% level, 95\% for corn and 92\% for soybeans.\footnote{This number is possibly under-estimated due to the \emph{undef} category, which corresponds to fields whose farm-equivalent coverage is identifiable only over an interval. Likely, some of these fields would have a value above 50\%, yet are not counted as $<50\%$ in this statistic. } Note that the 90\% and 95\% triggers are not offered by the RMA at the farm level, and we would expect that index insurance at 90\% does not perform better than farm-level insurance at the same 90\% trigger or even higher 95\%. Computing the \emph{farm-equivalent risk coverage} for the 90\% trigger is however informative on the randomness of our simulation: had we simulated data for a longer period than 30 years, the share of fields with index insurance better than the 90\% farm-level trigger would likely shrink. 

\subsection{Determinants of the utility of index insurance at the field level}

How are these metrics of the utility of index insurance linked to the measures of basis risk and to the fields's mean and variance? Our theory developed in Section~\ref{subsec:model} gave several predictions. Without surprise, the effect of basis risk is predicted to be negatively associated to the utility of index insurance.  The effect of the field's mean and variance is predicted to be respectively negative and positive when compared to no insurance, and both are ambiguous when compared to farm insurance. The empirical exercise here features, however, several departures from the model. First of all, we measure the benefit of index insurance using using the ratio of certainty equivalents, instead of the difference in utilities as in the theory. Second, the theoretical predictions are based on a second-order Taylor approximation, which ignores skewness and higher moments present in the data.\footnote{The raw data exhibits on average negative skewness.}

Table~\ref{tab:reg-EU2-vs-basis-plotLevel} shows a county-level fixed-effect regression of each metric on the basis risk measures and the mean and standard deviation of the yields. For the ``loss without indemnity'' measure of basis risk, we define now as loss those values below a threshold of 90\%. This is admittedly a high threshold (and higher than the 75\% threshold we used in Table~\ref{tab:sample-desk-stats}), but setting this value lower would entail discarding all fields whose relative minimum falls above the loss threshold. Looking first at the comparison of index insurance versus no insurance, we find that the two measures of basis risk have an opposite sign. The measure based on $1-R^2$ is negative as expected, indicating that fields with higher basis risk tend to have a lower utility from index insurance. However, the sign for the ``loss without indemnity'' basis risk measure is positive. This counter-intuitive result can possibly be understood keeping in mind the objections we raised above against the use of this measure. In a nutshell, it is merely a probability and hence not indicative of the extent of a possible damage, and is difficult to estimate in a short sample without making strong distributional assumptions. To compare the predictive power of each measure, we compute variable importance measures decomposing the $R^2$ into individual contributions using methods discussed and implemented in \citet{Groemping2006}. Based on  \citeauthor{Groemping2006}'s recommended \emph{lmg} method,\footnote{The \emph{lmg} measure solves the dependence on the ordering of the variables by computing all possible sequences of models and averaging over them.} our preferred measure of basis risk explains 43\% and 63\% of the total $R^2$ for corn and soybeans, while the alternative measure only explains 3\% and 5\% respectively. For further details, see Figure~\ref{fig:decomp_r2_relaimpo} in the appendix which shows the variable importance measures according to five alternative methods. Turning to the second utility metric of farm-equivalent risk coverage, we see now that   both measures of basis risk have a negative sign as expected. Once again, the variable importance measures indicate that our preferred measure of basis risk explains five to six times more of the total $R^2$ than the ``loss without indemnity'' probability.

\begin{table}
\caption{Determinants of utility of index insurance at the field level: fixed effects regression}
\begin{center}
\begin{threeparttable}
\begin{tabular}{l c c c c}
\toprule
 & \multicolumn{2}{c}{Corn} & \multicolumn{2}{c}{Soy} \\
\cmidrule(lr){2-3} \cmidrule(lr){4-5}
 & Area vs none & Area vs farm & Area vs none & Area vs farm \\
\midrule
Basis: $1-R^2$            & $-2.344^{***}$ & $-41.682^{***}$ & $-1.997^{***}$ & $-29.409^{***}$ \\
                          & $(0.122)$      & $(0.809)$       & $(0.075)$      & $(0.774)$       \\
Basis: $P(c>90\%|i<90\%)$ & $0.978^{***}$  & $-3.698^{***}$  & $0.434^{***}$  & $-3.534^{***}$  \\
                          & $(0.047)$      & $(0.118)$       & $(0.018)$      & $(0.175)$       \\
Field variance            & $0.001^{***}$  & $-0.007^{***}$  & $0.006^{***}$  & $-0.123^{***}$  \\
                          & $(0.000)$      & $(0.001)$       & $(0.001)$      & $(0.029)$       \\
Field mean                & $-0.016^{***}$ & $0.131^{***}$   & $-0.037^{***}$ & $0.506^{***}$   \\
                          & $(0.001)$      & $(0.006)$       & $(0.002)$      & $(0.015)$       \\
\midrule
Num. obs.                 & $1487774$      & $1469063$       & $1271190$      & $1235879$       \\
R$^2$ (full model)        & $0.590$        & $0.380$         & $0.576$        & $0.270$         \\
R$^2$ (proj model)        & $0.093$        & $0.327$         & $0.271$        & $0.220$         \\
Num. groups: county       & $597$          & $597$           & $597$          & $597$           \\
\bottomrule
\end{tabular}
\begin{tablenotes}[flushleft]
\scriptsize{\item $^{***}p<0.001$; $^{**}p<0.01$; $^{*}p<0.05$. \item Standard errors clustered at the county level. The regression is weighting fields by the number of years they were planted to corn or soybeans.}
\end{tablenotes}
\end{threeparttable}
\label{tab:reg-EU2-vs-basis-plotLevel}
\end{center}
\end{table}


The impact of the temporal variance on the utility of index insurance is very interesting. It has a positive impact on the utility of index insurance when assessed against no insurance, yet a negative impact when compared to farm insurance. The first one can be explained by the mechanical effect explained above: fields with higher variance are those that benefit most from any insurance. On the other hand, when comparing to farm insurance, fields with a higher variance tend to have a lower utility from index insurance. This suggests that the ``own protection'' effect from farm insurance (the first term in \ref{eq:Delta_area_vs_farm} is stronger than the mechanical ``any protection'' effect (the second term in \ref{eq:Delta_area_vs_farm}). In other words, the higher a field variance, the higher utility it will have from farm insurance, rendering the comparison with index insurance harder. 
Turning now to the effect of the field mean, we notice now a similar reversal as for the variance, albeit in the opposite direction: fields with higher mean tend to have a lower utility from index insurance when compared to no insurance, yet a higher when compared to farm insurance. This nicely reflects the prediction from the model. For the comparison to no insurance, the mean did not appear in equation \eqref{eq:Mir_varRed}, yet arose with a negative sign in the Taylor approximation. For the comparison with the farm insurance on the other side, the field mean has a positive sign in \eqref{eq:Delta_area_vs_farm}, which turns out to be stronger than the negative effect from the Taylor approximation.

Comparing finally the contribution of each of the four variables, Figure~\ref{fig:decomp_r2_relaimpo} reveals that the importance of the variables is flipped depending on the metric of index insurance used. For the comparison to no insurance, the field's mean and variance are the main regressors and basis risk measures play only a moderate role. On the other side,  for the comparison with farm insurance, our preferred measure of basis risk plays a large role, while the contribution of a field mean and variance is reduced to almost none.

Concluding these two sections dealing with data at the field level, our data and simulation each indicate that the area-based index insurance is rather effective. This is seen first from our measures of basis risk based on the raw dataset, and second from the expected utility measures obtained in the simulation. We find that index insurance provides at least a basic protection equivalent to 50\% of own protection for almost every field in the sample. Even when assessed against a much more stringent level of 85\%, we still find that for a good share of fields index insurance offers appropriate protection. We note several shortcomings of the popular definition of basis risk as ``probability of loss without indemnity'', and find that our preferred measure based on the $R^2$ from the field to county regression performs much better. Finally, we observe interesting diverging conclusions on the association between temporal variance and utility from index insurance: fields with higher variance tend to have higher utility from index insurance compared to no insurance, yet lower utility when assessed against farm insurance.

\subsection{Cross-county comparison}

\begin{figure}
\centering{}\includegraphics[width=0.95\columnwidth]{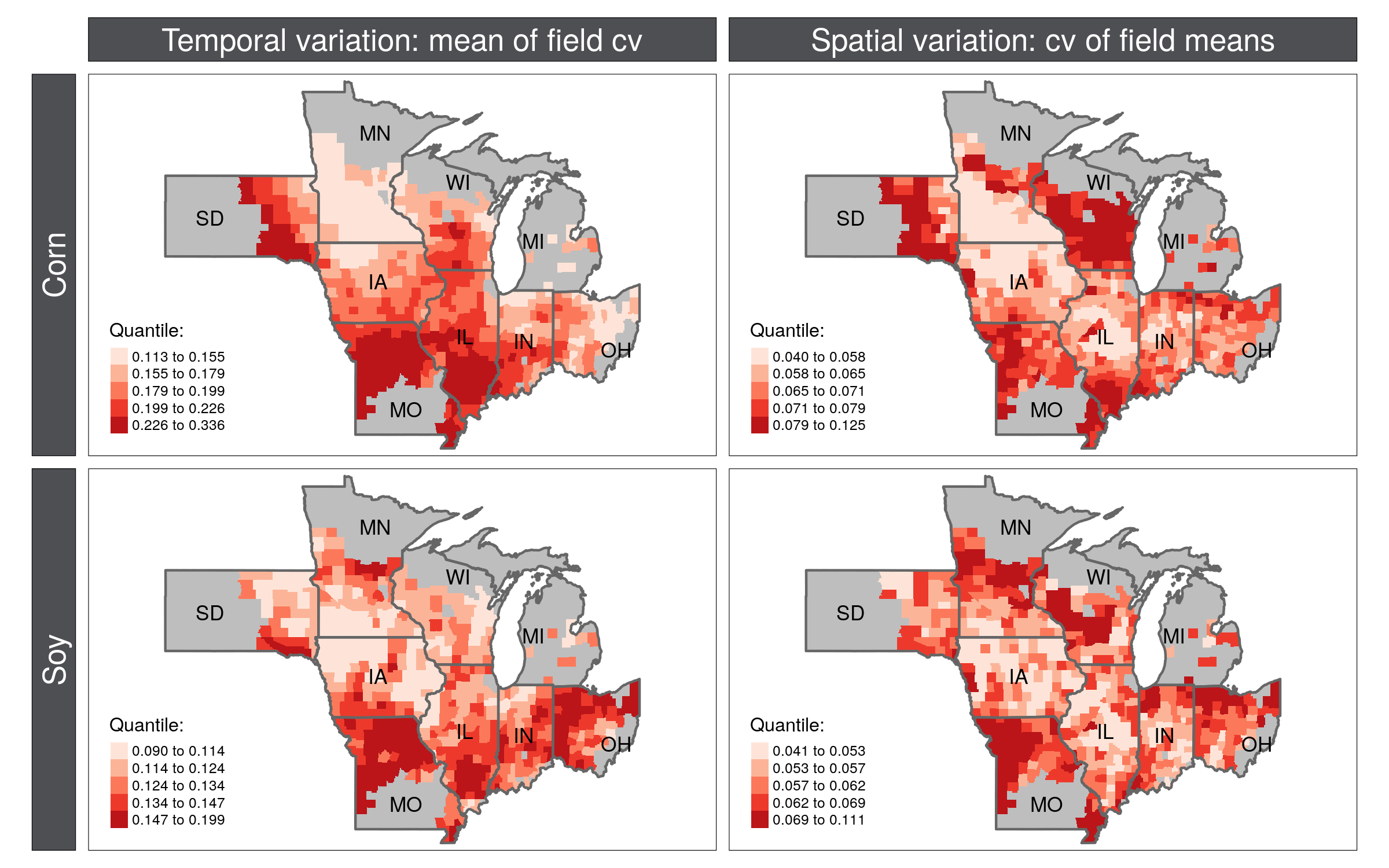}

\caption{County temporal and spatial variability\label{fig:Corn:-temporal-and-spat-var}}

\tablecomment{Source: raw data.}
\end{figure}

We proceed now to a cross-county comparison of the suitability of index insurance, relating our measures of index insurance utility to characteristics of the 597 counties in our dataset. We start by showing the spatial and temporal variation of yields between counties. The county-average \emph{temporal variability} is computed as the mean of every field's temporal variance. The county average \emph{spatial variability} is computed as the variance of the field means, and basically indicates how heterogeneous fields are within a county. These two measures are unique in the sense that they can only be obtained with field-level panel data. Our measure of average temporal variance for example is more informative than taking the variance of the county average (which could be obtained from official statistics), since the latter is a mix of temporal variance and spatial correlation.\footnote{Note indeed that the county average of field-level temporal variance is $\overline{\var(y_{i t})}=1/n\sum_i \sigma^2_i\equiv \overline{\sigma^2_i}$, while the variance of the county average, $\var(\bar{y}_{\cdot t})=1/N \overline{\sigma^2_i} + (N-1)/N \bar{\rho}$ is also a function of $\bar{\rho}$, the average covariance among fields.}
Figure~\ref{fig:Corn:-temporal-and-spat-var} shows these measures for each crop. The two measures show clear spatial patterns with a \emph{core versus periphery} pattern, where variability is relatively low in the center of the Corn Belt, in particular in Iowa (IA), North of Illinois (IL) and South of Minnesota (MN). On the other side, bordering regions such as South Dakota (SD), Missouri, and South of Illinois (IL), Indiana (IN), Ohio, Michigan (MI) and Wisconsin (WI), have markedly higher variability. This spatial pattern is similar across crops or variables, with a correlation between variables of 0.43 for corn and 0.36 for soy, while for the same variability measure, the correlation between corn and soy is 0.5 for the temporal variation, and 0.55 for the spatial variation. 

\begin{figure}
\centering{}

\includegraphics[width=0.95\columnwidth]{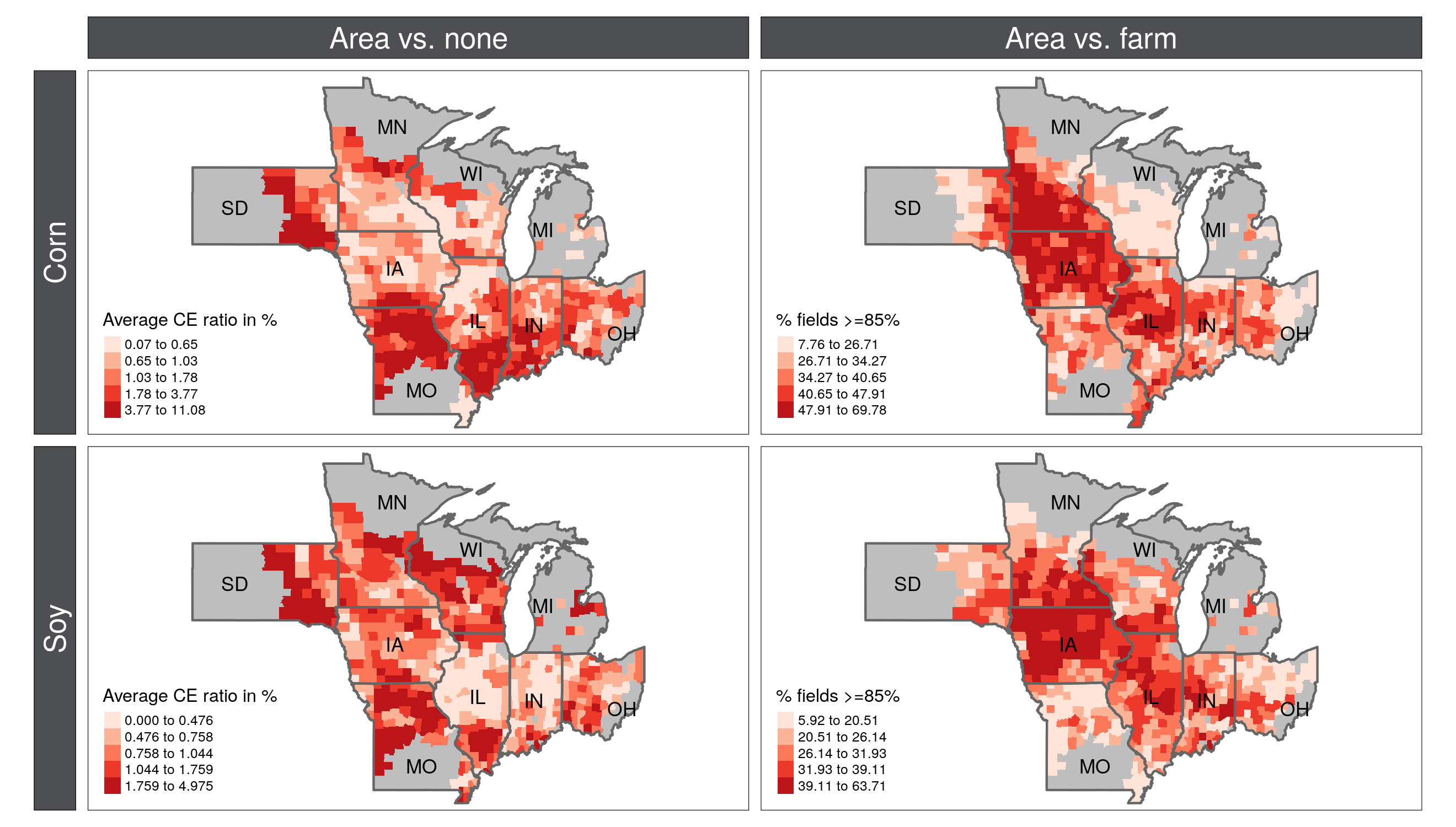}

\caption{Utility aggregated at the county level: comparing to no insurance and to farm insurance\label{fig:map-EU-versus-both}}

\tablecomment{Source: simulated data. }
\end{figure}

Turning now to the insurance utility measures based on the simulated data, the first column of Figure~\ref{fig:map-EU-versus-both} compares index insurance versus none by showing the percentage difference in certainty equivalent (CE) of index insurance compared to no insurance. Interestingly, the benefit of index insurance is high in the \emph{periphery} regions that have a high variability. It is indeed particularly high in Missouri (MO) and southern Illinois (IL), which are regions with both high temporal and spatial variability. In contrast, index insurance seems to be of more limited use in the \emph{core} regions such as northern Iowa (IA) and southern Minnesota. The second column of Figure~\ref{fig:map-EU-versus-both} compares on the other side the index insurance to the farm-based insurance, using our measure of \emph{farm-equivalent risk coverage} at 85\%. We aggregate this measure by counting the number of fields within each county for which index insurance is at least as good as a 85\% farm insurance. The conclusion is now reversed: counties in the \emph{periphery} have fewer fields with a 85\% farm-equivalent risk coverage, while those in the \emph{core} show much higher benefits from index insurance.  Most counties in Iowa (IA) have 40\% or more of their fields which would benefit from index insurance even compared to farm-based protection at 85\%.

\begin{figure}

\centering{}\includegraphics[width=0.95\columnwidth]{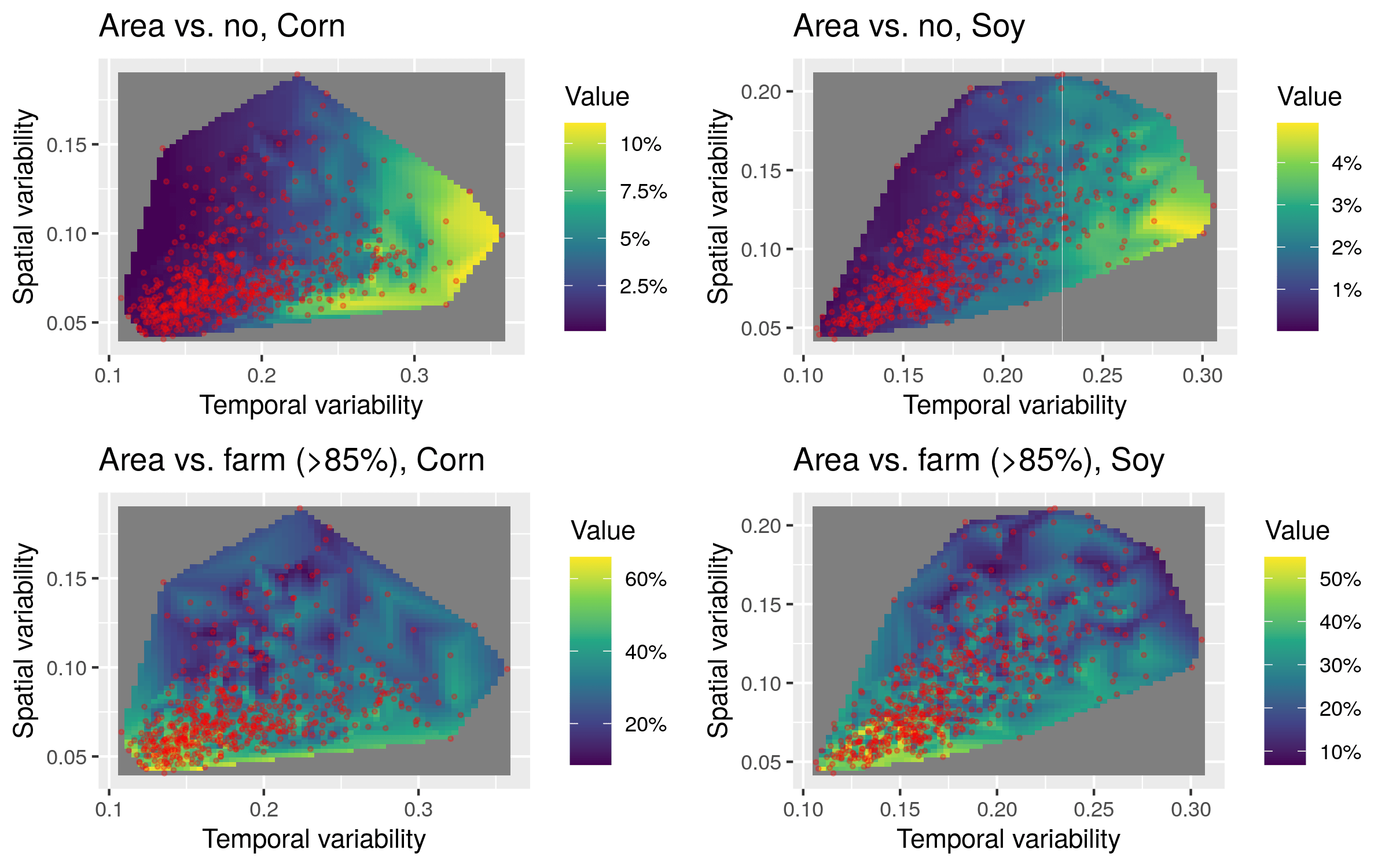}

\caption{Heatmap of utility of insurance according to county temporal and spatial variability\label{fig:heatmap-2Var}}

\tablecomment{Each panel shows a heatmap of the utility of index insurance. The heatmap was obtained by interpolating the utility values between the actual locations of the counties represented with red dots. }

\end{figure}

The previous results reveal an interesting reversal: counties where index insurance appears to be the most useful when assessed against no insurance turn out to be counties in which index insurance is the least beneficial in terms of farm-equivalent protection. This paradox stems from the fact that temporal and spatial variability happen to be positively correlated, yet these have opposite effects on the usefulness of index insurance. To make this point clearer, Figure~\ref{fig:heatmap-2Var} shows a heat map of the utility of index insurance (compared to either no insurance or to farm insurance), projected in the spatial-temporal variability space. The heat map is obtained by interpolating the utility metrics based on the actual values of the counties, whose location is shown with red dots. The first row shows the values for the utility of index insurance versus no insurance. The highest utility is found on the east side of the space, where temporal variability is highest. The gradient of utility along the temporal variability (x-axis) is so strong that it appears to obscure the effect of spatial variability (y axis), which seems to have almost no impact on the utility of index insurance. However, when looking now at the utility of index insurance compared to farm insurance on the second row, results are reversed, much as we saw from the maps in Figure~\ref{fig:map-EU-versus-both}. The highest utility of index insurance is now on the south-west part of the graph, where spatial and temporal variability are lowest. The east portion that was previously giving the highest utility when compared to no insurance gives now almost the lowest utility.

To confirm the results from the graphical analysis, we run a regression of the two utility metrics on county's average spatial variance, temporal variance, $R^2$ and county mean. This regression is similar to the regression at the field-level in Table~\ref{tab:reg-EU2-vs-basis-plotLevel}, the main difference being now that we are dealing with county-aggregated values. As our focus is on comparing counties, we run this cross-section regression without any regional fixed effects. Results shown in Table~\ref{tab:reg-EU2-vs-Sstats}  confirm the reversal in sign for the temporal variance: the coefficient is positive for the utility of index insurance versus none, yet negative when compared to farm insurance. On the other hand, the coefficients for $R^2$ and the spatial variance are constant across utility metrics: counties with higher $R^2$ or lower spatial variance have higher utility from index insurance according to each metric. Except for the in-significant effect of the county average, the results at the county level are much in line with those from the field-level regression shown earlier in Table~\ref{tab:reg-EU2-vs-basis-plotLevel}. We observe at each level the reversal of the temporal variance based on the metric of utility of index insurance. We observe also in each case that the $R^2$ measure has a stronger explanatory power when comparing index insurance to farm insurance rather than to no insurance (see relative importance measures in Table~\ref{tab:relimpCounty}).

\begin{table}
\caption{Determinants of utility of index insurance at the county level}
\begin{center}
\begin{tabular}{l c c c c}
\toprule
 & \multicolumn{2}{c}{Corn} & \multicolumn{2}{c}{Soy} \\
\cmidrule(lr){2-3} \cmidrule(lr){4-5}
 & Area vs. none & Area vs. farm & Area vs. none & Area vs. farm \\
\midrule
$R^2$ county      & $0.10^{***}$  & $0.37^{***}$  & $0.09^{***}$  & $0.29^{***}$  \\
                  & $(0.02)$      & $(0.03)$      & $(0.02)$      & $(0.04)$      \\
Temporal Variance & $0.95^{***}$  & $-0.30^{***}$ & $1.09^{***}$  & $-0.24^{***}$ \\
                  & $(0.02)$      & $(0.03)$      & $(0.03)$      & $(0.05)$      \\
Spatial Variance  & $-0.20^{***}$ & $-0.39^{***}$ & $-0.30^{***}$ & $-0.32^{***}$ \\
                  & $(0.02)$      & $(0.04)$      & $(0.03)$      & $(0.05)$      \\
County Mean       & $0.00$        & $0.00$        & $0.00$        & $0.00$        \\
                  & $(0.00)$      & $(0.00)$      & $(0.00)$      & $(0.00)$      \\
\midrule
R$^2$             & $0.84$        & $0.59$        & $0.82$        & $0.46$        \\
Num. obs.         & $597$         & $597$         & $597$         & $597$         \\
\bottomrule
\multicolumn{5}{l}{\scriptsize{Coefficients are standardized by re-scaling variables, hence no intercept is included.}}
\end{tabular}
\label{tab:reg-EU2-vs-Sstats}
\end{center}
\end{table}

The reversal we document here leads to an interesting dilemma: those places where risk is highest and hence where insurance is the most needed are also those where index insurance is the least useful. Said differently, index insurance leads to good farm-equivalent coverage only in those counties that have the lowest risk. Because of the positive correlation between temporal and spatial variability, when average individual risk increases, so does the spatial variability, deteriorating the benefits of index insurance. A direct consequence of this is that selecting good zones for index insurance is a difficult task: for one, easy available statistics such as the temporal variance of the zone average\footnote{Remember that our measure of temporal variability used here is derived from the average field-level variance, which is not equal to the variance of the average.} are potentially misleading, leading to choose zones where index insurance offers the lowest farm-equivalent coverage. What is clearly needed beyond the variance of the county average is information on the spatial variability and spatial correlation, which are much harder to obtain in practice.

\section{Robustness checks\label{sec:Robustness}}

In this section, we carry multiple robustness checks. We look first at the effect of using raw data instead of the simulated data in \ref{subsec:robust_simulated}. In \ref{subsec:robust_sample_size} we investigate the effect of changing the sample size, while in \ref{subsec:robust_subsidies} we include subsidies. Finally, we use cumulative prospect theory instead of expected utility in \ref{subsec:robust_prospect}.

\subsection{Robustness checks: effect of using simulated data}
\label{subsec:robust_simulated}

Our analysis so far relied on the simulated data. The main reason was that the raw data has a large amount of missing values due to the practice of crop rotation. For a given field, we typically only observe 50\% of the corn and 50\% of the soy yields over a given period. This implies that even if the premiums are fair at the county level (i.e. the premium is equal to the average of the indemnities), they might be very unfair or very favorable for a given field depending on the cropping sequence. Imagine indeed that a field cultivated corn only in the drought year 2012, receiving a huge indemnity yet paying a small premium once. On the other hand, a field might cultivate corn every year but 2012, paying a high premium every year yet not receiving the large indemnity for 2012. Using the simulated data allowed us to avoid this randomness, by filling-in missing years. A second reason for using the simulated data was that it permits to extend the sample over time, and usually also to extend the empirical support of the yield distribution. This is particularly useful given that our measure of farm-equivalent risk coverage is not defined beyond the empirical minimum.\footnote{Remember that if the minimum of a field is say 80\% of its mean, the utility of a coverage at 75\% is undefined.} Extending the range of the data hence reduces the number of undefined cases. 

We investigate now the effect of using the raw data instead of the simulated one. We do this in two steps. In the first step, we isolate the effect of the possible \emph{unfairness} of the county premium, and construct premiums that are fair at the field level. That is, we assume the insurer is computing a field-specific premium even for the county insurance, taking into account only the years in which a field is planted to the specific crop. In a second step, we relax that assumption, and use premiums that are fair at the county level only (and hence potentially unfair or very favorable).

\begin{figure}
\centering{}
\includegraphics[width=1\columnwidth]{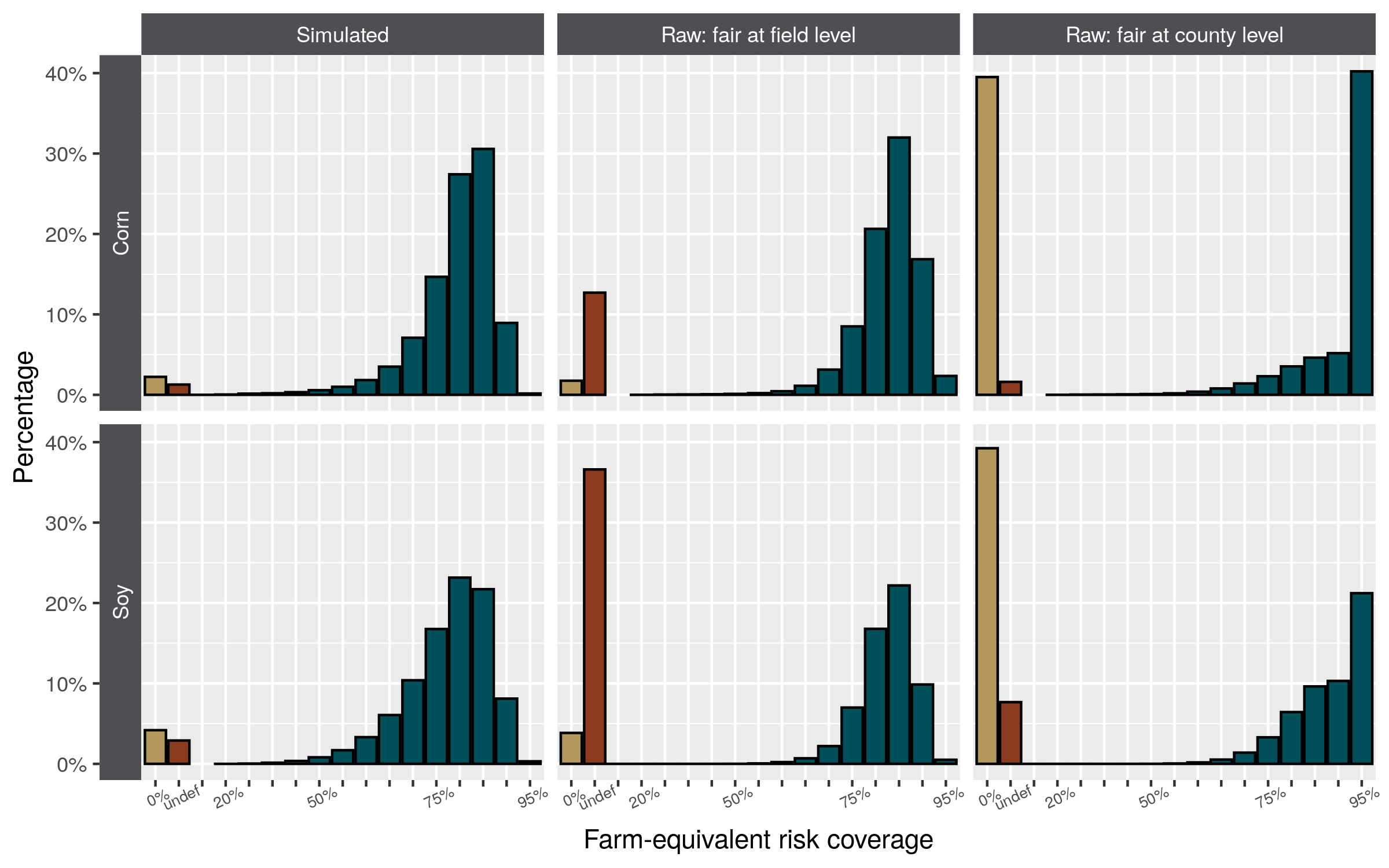}

\caption{Effect of using simulated versus raw data \label{fig:histo_FImax_simul_vs_raw}}

\end{figure}

Figure~\ref{fig:histo_FImax_simul_vs_raw} shows the histogram of our \emph{farm-equivalent risk coverage} metric, using 1) the simulated data (hence reproducing Figure~\ref{fig:Comparing-CE_vs_farm} above), 2) the raw data with county premiums that are fair at the field level and 3) the raw data with county premiums fair at the county level. Looking first at the raw data with premiums fair at the field level (second column), we see that the main difference lies in the \emph{undef} category, which corresponds to fields for which index insurance was higher than no insurance (0\%), yet lower than the level of the lowest-observed farm insurance, and hence is not clearly defined. Abstracting from this category, results look qualitatively similar. Turning to the last column  showing fields with county-fair premiums, we see a striking increase in fields that either have a 0\% equivalent coverage, or 100\%. This illustrates well the problem of premiums being ``over-'' and ``under-fair'' premiums depending on whether or not a field planted the crop in the bad year 2012. Indeed for corn, among those fields for whom we find a 0\% farm-equivalent value, 92\% of those did not plant to corn that year, while among those that have the maximum 95\% farm-equivalent value, 95\% planted corn in 2012. For soybeans, the influence of the year 2012 is less strong, which is partly explained by the fact that the impact of the 2012 drought on yields was lower compared to corn.

\subsection{Robustness check: simulating using longer sample sizes}\label{subsec:robust_sample_size}

In the main analysis we used simulated data over a period of 29 years. Compared to other studies, this is rather a small number: \citet{YeHuEtAl2020} use for example 1000 simulations for each farm. Our choice for 29 years is the result of a trade-off between computational burden on one side and accuracy of the risk calculation on the other side. Noting that our sample has more than 1.8 million (M) fields and 2.8M field-crop pairs, using 29 years of data already results into a dataset of more than 80M rows. 

We investigate here nevertheless the impact of increasing the sample size of the simulated yields series. Note that the initial simulated data was simulating using (detrended) NASS county aggregates from 1990 to 2018 plugged into the field-county regression (\ref{eq:MirReg}). This approach cannot be used anymore for larger series, so we proceed instead to simulating NASS data itself. To do so, we estimate an AR(2) model on each individual series, and predict from this model. Innovations for the predictions are taken from the empirical distribution of the residuals of the fitted AR model. We opt for the empirical distribution for the innovations as it allows to capture large negative shocks that would be difficult to model with standard parametric distributions.

\begin{figure}
\centering{}
\includegraphics[width=0.5\columnwidth]{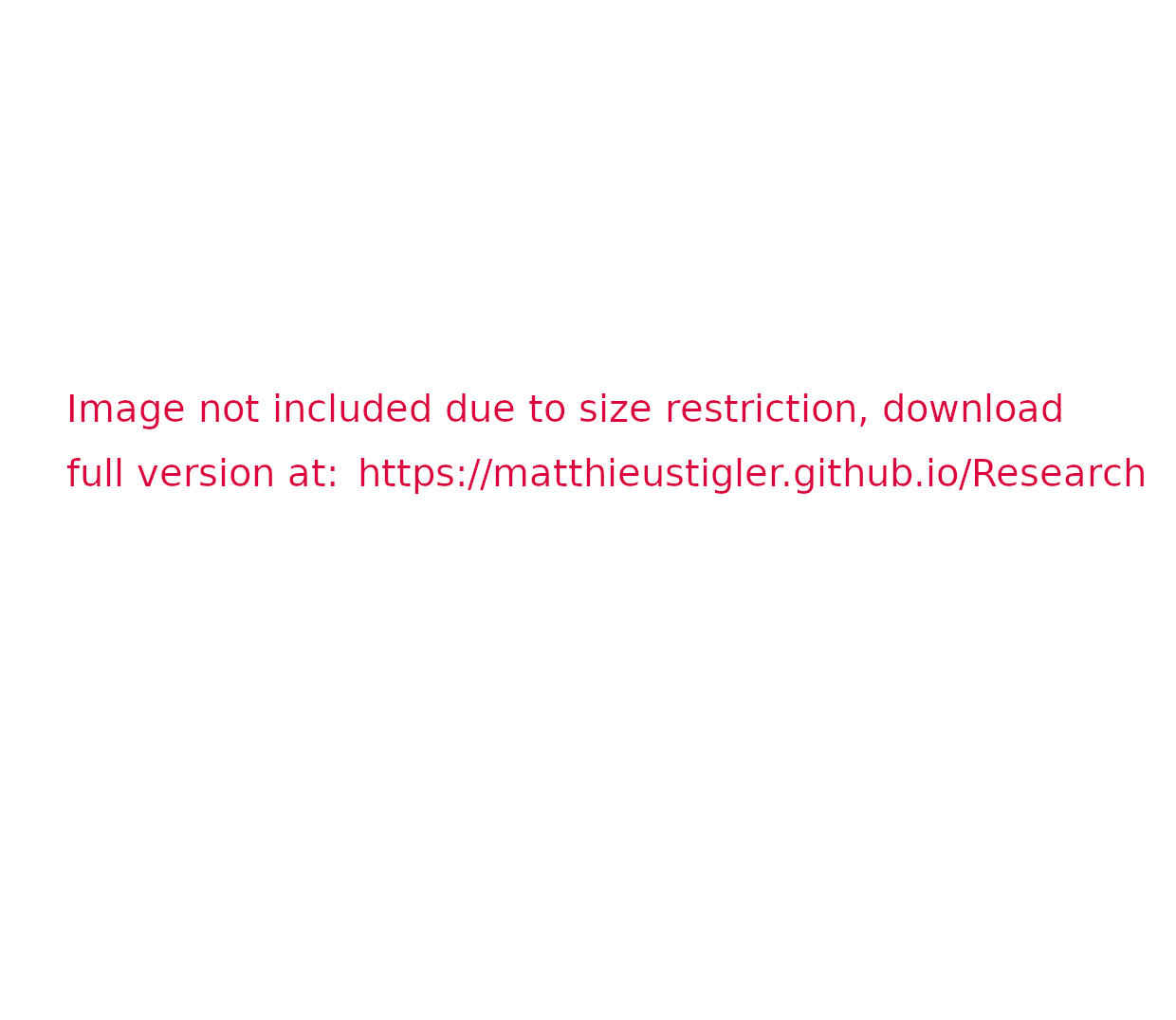}

\caption{Effect of changing the sample size \label{fig:robust_sample_size}}

\end{figure}

Figure~\ref{fig:robust_sample_size} shows the county-level measure of the percentage of fields with index insurance at least as good as 85\%, as well as 90\%. The red line indicates the average over all the counties. Blue dots represent the sample sizes taken into consideration: 20, 30, 60, 100, 250 and 500. Results indicate that with larger sample sizes, our measure of the benefits of index insurance decreases, yet get stabilized relatively quickly for sample sizes of 100 or larger. This suggests that there is a small upward bias in our estimates on the order of 5\%. The second panel shows the results for the percentage of fields for which index insurance is at least as good as a 90\% farm-insurance. Remember that the coverage for the index insurance has been selected at 90\% throughout this paper. Intuitively, we would expect that at equal coverage, farm-level insurance does better than area-based. Presence of fields for which area schemes are nevertheless better than farm-based was assumed so far to arise from simulation noise. Figure~\ref{fig:robust_sample_size} partly confirms this intuition: the percentage decreases with bigger sample size. However, it does not goes to zero, indicating the presence of fields which constantly prefer area-based insurance.

\subsection{Robustness check: taking subsidies into consideration}\label{subsec:robust_subsidies}

In the main analysis, we considered fair premiums. However, premiums are actually heavily subsidized by the Federal Government. Subsidies vary depending on the level of coverage and the type of scheme, see Table~\ref{tab:Subsidy-rate-for-area-farm}. The subsidies start at around 60\% for the lowest level of coverage, and decrease to 40-50\% for the highest ones. Subsidies are higher for the area-based scheme, in particular for higher coverage levels. This suggests that taking subsidies into account should increase the attractiveness of area-based insurance  compared to our benchmark.

To verify the impact of subsidies, we re-run the same analysis as above, this time applying the subsidies to the premiums. Without surprise, our two metrics of the utility of index insurance both increase with subsidies for almost all fields ($>99\%$). As a consequence, the aggregate number of fields preferring index insurance to no insurance or to the highest farm-level insurance also increase. 


\subsection{Robustness check: using cumulative prospect theory}\label{subsec:robust_prospect}

The analysis so far was based on the expected utility (EU) framework. But several authors have pointed out that the widespread under-coverage observed in practice cannot be explained by expected utility \citep{Babcock2015,FengDuEtAl2020}. According to EU theory, farmers should seek maximal coverage with fair premiums. 
\citet{DuFengEtAl2017} develop a framework to include subsidies in expected utility computations, yet find that this still does not explain the low coverage chosen in practice. \citet{Babcock2015} uses instead cumulative prospect theory (CPT), finding that it captures better the observed behavior of the three farms he considers. Cumulative prospect theory \citep{TverskyKahneman1992} allows to capture phenomena like loss aversion, probability weighting and reference dependence. Reference dependence refers to the existence of a \emph{reference} point below which outcomes are considered losses, and above which values are considered gains. It is not obvious what this reference point should be for farmers' choice of crop insurance. We follow here \citet{Babcock2015}, and consider two possible reference points.\footnote{We do not consider his third reference point which is based on indemnities only. This is because with fair premiums, considering indemnities on their own amounts to choosing a lottery that increases risk yet has zero expected gain. No farmers would want such a lottery.} The first one includes the expected yield plus the premium. The second uses only the expected yield, considering that premiums are considered a sunk cost. For the choice of the value and decision weighting functions, we use exactly the same functions and parameters as in \citet{Babcock2015}, who used values directly derived from \citet{TverskyKahneman1992}. Like \citet{Babcock2015}, we use the empirical distribution and hence assign weights 1/T to each yield outcome. 

\begin{figure}
\centering{}
\includegraphics[width=1\columnwidth]{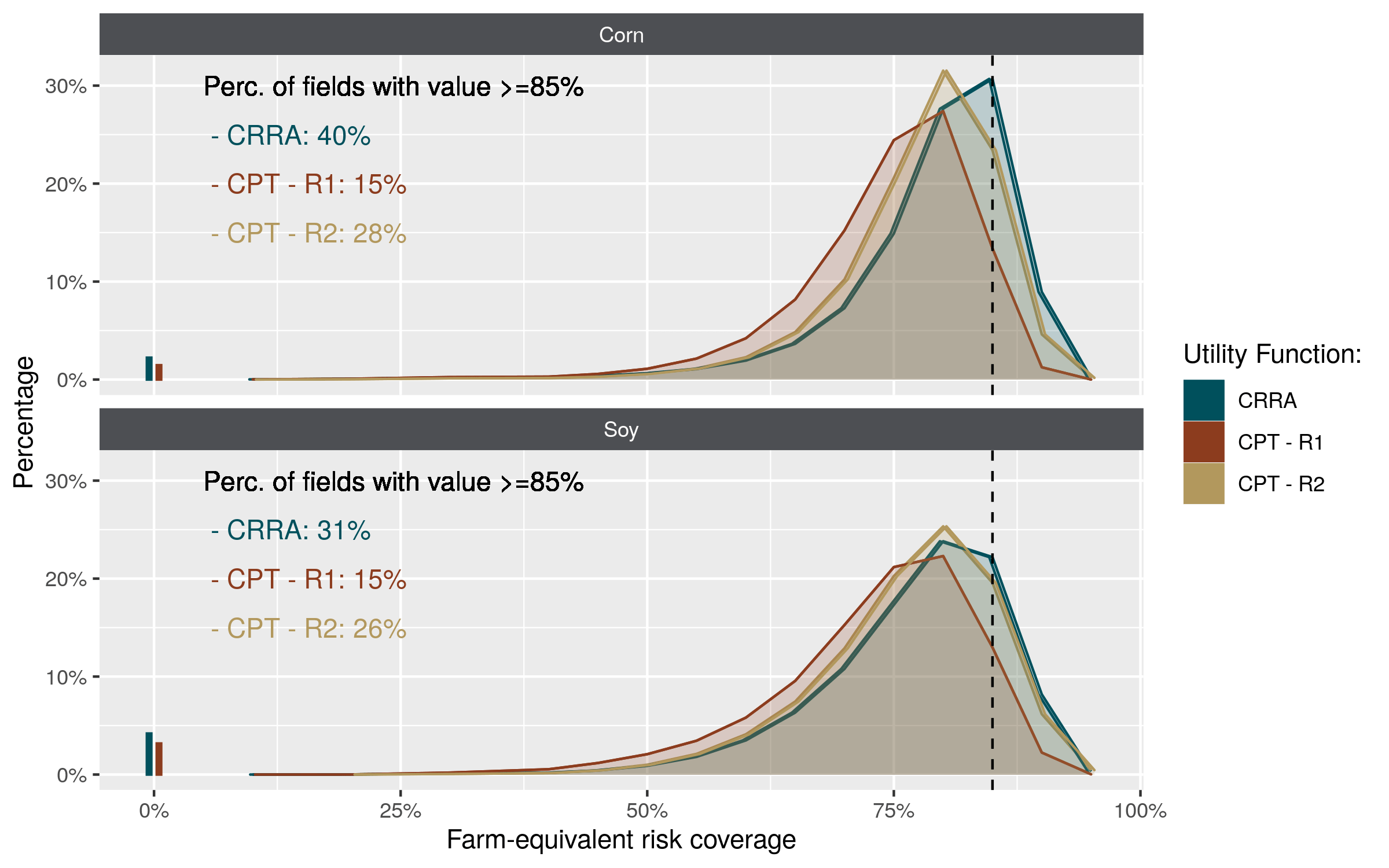}
\caption{Effect of using cumulative prospect theory \label{fig:robust_CPT}}
\end{figure}

We re-run the analysis evaluating now our metrics of index insurance with 
the cumulative prospect theory functions. Figure~\ref{fig:robust_CPT} shows the distribution of the farm-equivalent coverage with the standard expected utility (CRRA), as well as the CPT with the two reference points.  \emph{R1} refers to the point including expected yields plus premium and \emph{R2} to expected yields only. Using CPT induces a reduction of the benefits of index insurance, in particular using R1. The percentage of fields who prefer index insurance to farm insurance at 85\% falls from a 30\% to 40\% under expected utility to a 15\% under CPT R1, and to 26-28\% under the R2 target. This is definitely an interesting results, as it seems more realistic than the 30\% to 40\% predicted under expected utility. Nevertheless, that number is still quite larger than the take-up observed in practice, which is never more than 5\%.

\section{Conclusion}

In this study, we investigate the suitability of crop insurance in the US using a unique dataset of nearly two million fields observed over 20 years through satellite remote sensing. We run a large-scale simulation seeking to replicate observed yields as closely as possible, yet abstracting from moral hazard or adverse selection issues. We develop several metrics of suitability of index insurance based on expected utility theory, comparing index insurance to no insurance but also to farm-level insurance. Thanks to the very large scale of our dataset spanning close to 600 index insurance zones, we are also able to investigate the characteristics of the counties which make insurance more beneficial. 


Our first contribution is to show that index insurance performs surprisingly well, shedding a new positive light for index insurance. Our simulations show that absent adverse selection and moral hazard, index insurance brings a positive improvement for almost all fields. When expressed in our new measure of farm-equivalent coverage, index insurance is at least as good as a 50\% farm coverage for a majority of fields,  indicating that it can serve the basic function of protecting against catastrophic events. Furthermore, when assessed against the highest-available level of 85\%, 30\% of the fields still benefit more from an index insurance at the 90\% coverage level. Our results are robust to relaxing several assumptions of the model, although they tend altogether to slightly reduce the benefits observed. The largest 
changes stem from using cumulative prospect theory instead of expected utility. This is an interesting result that will deserve further analysis. 

Our second contribution is to investigate the spatial determinants of the suitability of index insurance, and we uncover an interesting paradox. We start by developing a formal theory of the benefits of index insurance compared to farm insurance, extending the theory of \citet{Miranda1991}. Our key contribution is to show that the metric chosen to assess those benefits plays a crucial role, and seemingly similar metrics can lead to opposite results. When assessed against no insurance, index insurance seems to be the most beneficial in the counties in the outer Corn Belt, which have a higher temporal variability. On the other hand, when assessed against our new measure of farm-equivalent risk, index insurance is now the most beneficial in the counties in the core of the Corn Belt, which have the lowest temporal variability. This result is explained by the fact that temporal and spatial variability tend to be correlated at the county level. While temporal variance increases the benefit of index insurance (as it does for any insurance), spatial variability reduces it. This result has important practical implications for the design of index insurance, highlighting the sensitivity of the choice of metric, and the unreliability of an analysis solely based on the most easily available statistic, the temporal variance of (average) yields.


This study could be extended in several ways. For one, we assumed away adverse selection and moral hazard, and relaxing each of these assumptions would be interesting on its own. We ruled out adverse selection by calculating fair premiums ex-post, implying that risk is perfectly measured. Predicting ex-ante premiums, following the large literature based on \cite{HarriCobleEtAl2011}, would be a worthwhile extension, opening the door to models of adverse selection following \citet{JustCalvinEtAl1999}. Our results showing the different predictions obtained from cumulative prospect theory are also very promising. Those could be extended to the question of the coverage of farm-level insurance, extending the work by \citet{Babcock2015}. Finally, multiple improvements could  be done on the methodological side when we seek to model the correlation between yields. This is definitely a high-dimension problem given that we have at most twenty time periods yet hundreds or even a few thousand of variables. While there exist several techniques to model covariance matrices in very large dimension, little guidance is available to address the case with missing values often encountered with yield data.


\clearpage
\appendix

\setcounter{table}{0}
\setcounter{figure}{0}
\renewcommand{\thetable}{A.\arabic{table}}
\renewcommand\thefigure{A.\arabic{figure}}    

\section{Appendix}

\subsection{Proofs}

\begin{prop}[Second-order approximation of expected utility] 
\label{th:taylor}
$\E[u(y^A)]-\E[u(y^B)]\approx1/2 u''(\mu)\left(\sigma^2_{y^A}-\sigma^2_{y^B}\right)$
\end{prop}

\begin{proof}
Use first a second-order Taylor expansion around the mean $\mu$: $\E[u(y)]\approx u(\mu) + u'(\mu)\E[y-\mu]+1/2 u''(\mu)\E(y-\mu)^2=u(\mu) +1/2 u''(\mu)\sigma^2_y$. The second term $\E[x-\mu]$ equals zero by definition of $\mu=\E[x]$, while the third term corresponds to the variance $\sigma^2_y\equiv\E(y-\mu)^2$

Now whenever we compare two schemes with fair premiums ($\mu_A=\mu_B$), comparing the expected utility amounts to comparing : $\E[u(y^A)]-\E[u(y^B)]=u(\mu^A) +1/2 u''(\mu^A)\sigma^2_{y^A}-u(\mu^B) -1/2 u''(\mu_B)\sigma^2_{y^B}=1/2 u''(\mu)(\sigma^2_{y^A}-\sigma^2_{y^B})$

\end{proof}

\subsection{Supplementary figures}

\begin{figure}
\includegraphics[width=0.95\columnwidth]{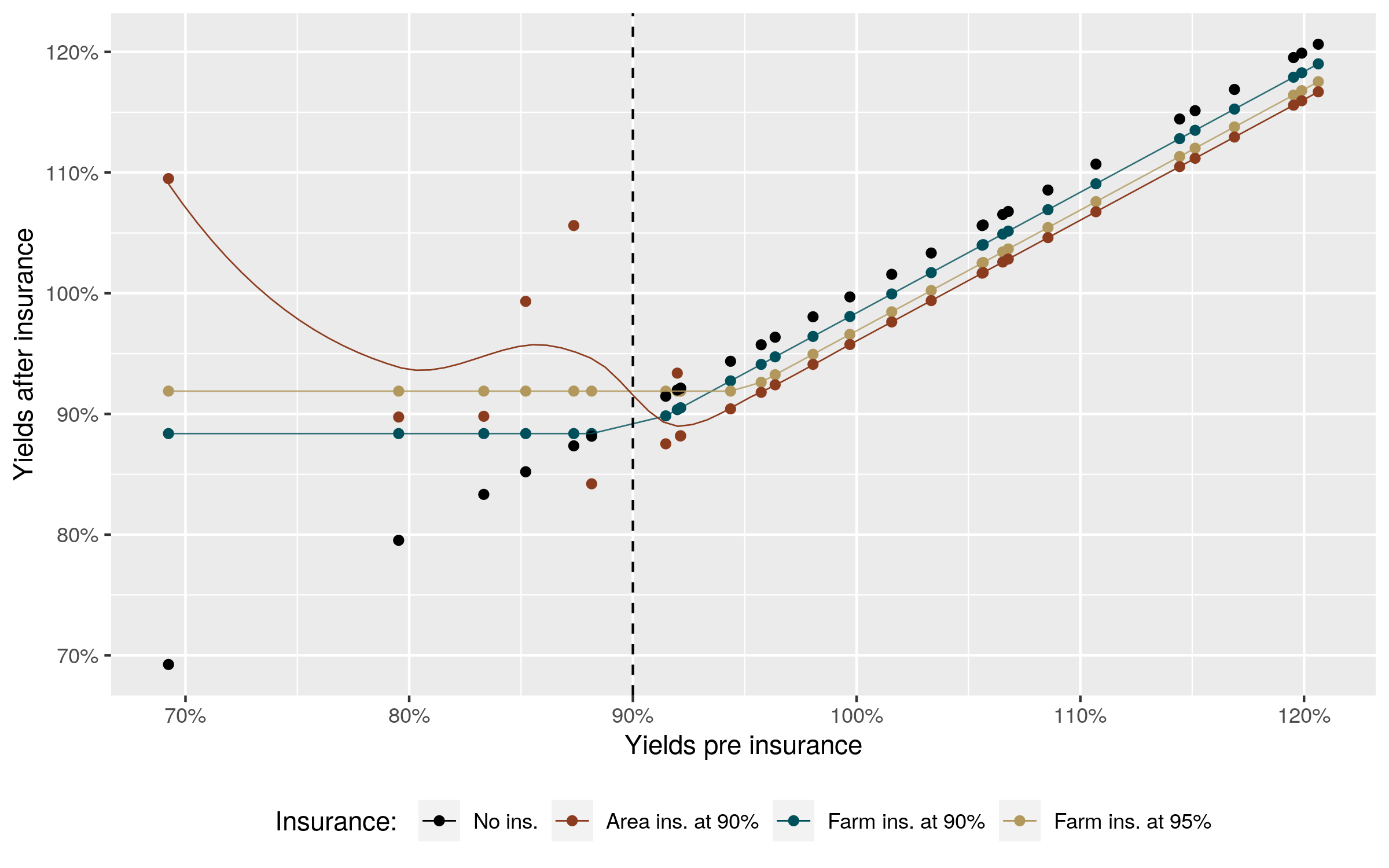}

\caption{Yield with and without insurance: illustration for a single field\label{fig:yields_pre_post_1plot}}

\end{figure}

\begin{figure}
\includegraphics[width=0.95\columnwidth]{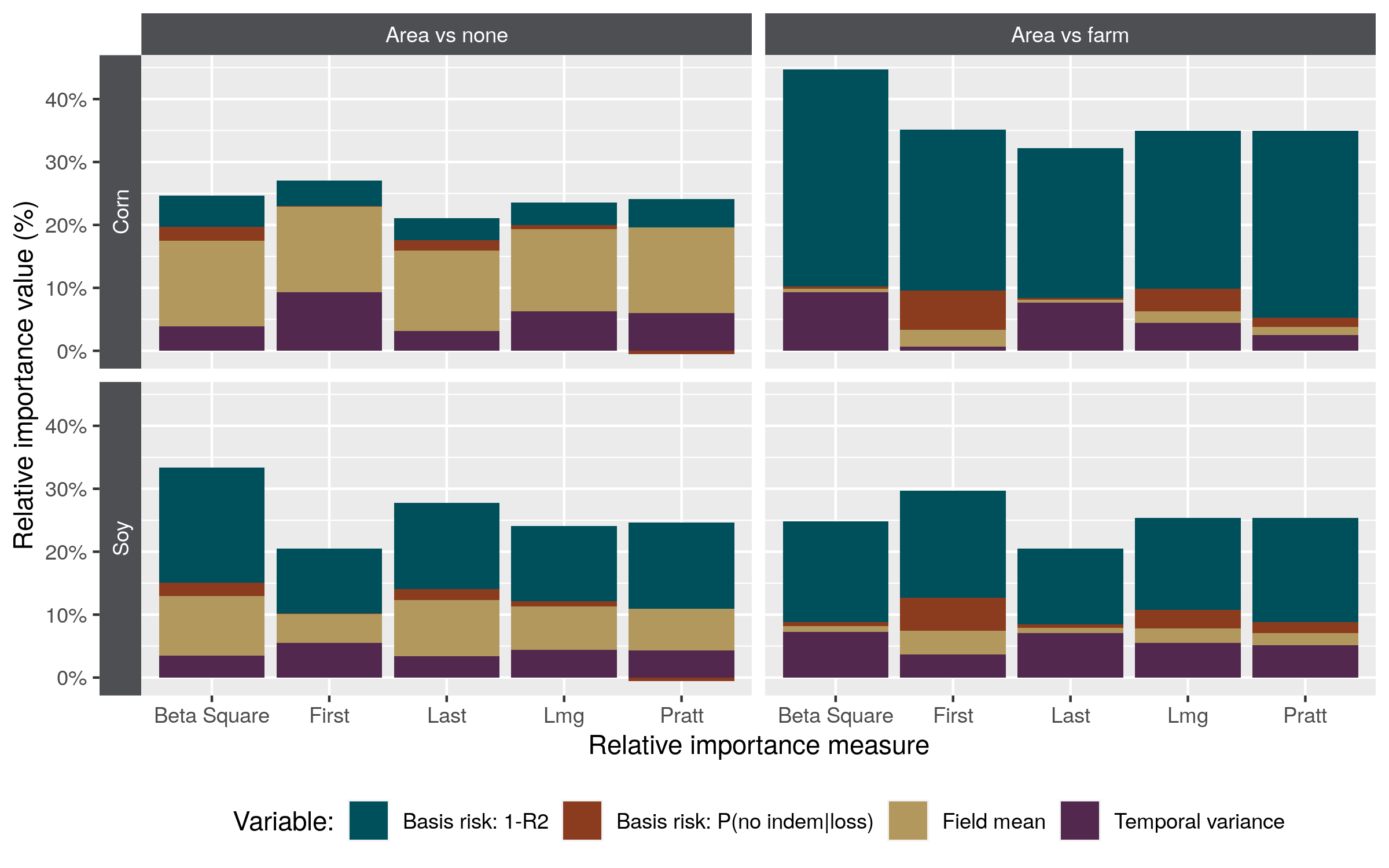}

\caption{Relative importance measures for the variables in the field-level regression \label{fig:decomp_r2_relaimpo}}

\end{figure}

\begin{table}[ht]
\centering
\caption{Relative importance measures for the county-level regression} 
\label{tab:relimpCounty}
\begin{tabular}{llrrrr}
  \toprule
 &&\multicolumn{2}{c}{Area vs. none} &\multicolumn{2}{c}{Area vs. farm}\\
\cmidrule(l){3-4}\cmidrule(l){5-6}
Crop & Variable & Value  & Percentage  & Value  & Percentage  \\ 
  \midrule
Corn & County Mean & 0.20 & 23.05 & 0.08 & 12.79 \\ 
   & $R^2$ county & 0.06 & 7.19 & 0.21 & 33.86 \\ 
   & Spatial Variance & 0.03 & 3.66 & 0.26 & 41.87 \\ 
   & Temporal Variance & 0.56 & 66.10 & 0.07 & 11.48 \\ 
  Soy & County Mean & 0.11 & 13.76 & 0.11 & 23.76 \\ 
   & $R^2$ county & 0.04 & 5.06 & 0.11 & 22.27 \\ 
   & Spatial Variance & 0.10 & 12.39 & 0.17 & 35.31 \\ 
   & Temporal Variance & 0.57 & 68.79 & 0.09 & 18.66 \\ 
   \bottomrule
\multicolumn{6}{l}{Relative importance computed with the lmg metric from R package relimp}\\
\end{tabular}
\end{table}

\begin{table}[ht]
\centering
\caption{Estimates of basis risk using the \emph{loss without indemnity} definition} 
\label{tab:basis-risk-by-level}
\begin{tabular}{lllll}
  \toprule
Crop & Level x & P(field $<$ x) & P(county$>$90\%$|$field$<$x) & Confidence Interval (95\%) \\ 
  \midrule
Corn & 90\% & 24.67\% & 30.48\% & [30.43 - 30.53] \\ 
   & 85\% & 16.49\% & 18.72\% & [18.67 - 18.77] \\ 
   & 75\% & 7.52\% & 6.7\% & [ 6.66 -  6.75] \\ 
   & 65\% & 4.24\% & 2.87\% & [ 2.83 -  2.91] \\ 
   & 55\% & 2.65\% & 1.79\% & [ 1.76 -  1.83] \\ 
   & 50\% & 2.09\% & 1.56\% & [ 1.52 -  1.60] \\ 
  Soy & 90\% & 21.96\% & 46.34\% & [46.28 - 46.41] \\ 
   & 85\% & 12.79\% & 35.63\% & [35.55 - 35.71] \\ 
   & 75\% & 3.19\% & 24.54\% & [24.40 - 24.67] \\ 
   & 65\% & 0.44\% & 31.52\% & [31.14 - 31.90] \\ 
   & 55\% & 0.03\% & 40.55\% & [38.97 - 42.13] \\ 
   & 50\% & 0.01\% & 29.87\% & [26.89 - 32.85] \\ 
   \bottomrule
\multicolumn{4}{l}{Source: raw dataset.}\\
\end{tabular}
\end{table}

\begin{table}
\caption{Subsidy rate for farm- and area-based plans, yield protection\label{tab:Subsidy-rate-for-area-farm}}

\begin{centering}
\begin{tabular}{cccc}
\toprule 
 &  & \multicolumn{2}{c}{Subsidy rate}\tabularnewline
\midrule 
Coverage type & Coverage Level & Farm yield & Area yield\tabularnewline
\midrule
\midrule 
Catastrophic & 50\% & 100\% & \tabularnewline
\midrule 
\multirow{9}{*}{Additional} & 50\% & 67\% & \tabularnewline
\cmidrule{2-4} \cmidrule{3-4} \cmidrule{4-4} 
 & 55\% & 64\% & \tabularnewline
\cmidrule{2-4} \cmidrule{3-4} \cmidrule{4-4} 
 & 60\% & 64\% & \tabularnewline
\cmidrule{2-4} \cmidrule{3-4} \cmidrule{4-4} 
 & 65\% & 59\% & \tabularnewline
\cmidrule{2-4} \cmidrule{3-4} \cmidrule{4-4} 
 & 70\% & 59\% & 59\%\tabularnewline
\cmidrule{2-4} \cmidrule{3-4} \cmidrule{4-4} 
 & 75\% & 55\% & 59\%\tabularnewline
\cmidrule{2-4} \cmidrule{3-4} \cmidrule{4-4} 
 & 80\% & 48\% & 55\%\tabularnewline
\cmidrule{2-4} \cmidrule{3-4} \cmidrule{4-4} 
 & 85\% & 38\% & 55\%\tabularnewline
\cmidrule{2-4} \cmidrule{3-4} \cmidrule{4-4} 
 & 90\% & - & 51\%\tabularnewline
\bottomrule
\end{tabular}
\par\end{centering}

\vspace{0.3cm}
\tablesource{RMA Insurance Handbook}
\end{table}

 \clearpage 

\bibliographystyle{ecta}
\bibliography{BiblioSynced}

\begin{thebibliography}{52}
\newcommand{\enquote}[1]{``#1''}
\expandafter\ifx\csname natexlab\endcsname\relax\def\natexlab#1{#1}\fi

\bibitem[\protect\citeauthoryear{Babcock}{Babcock}{2015}]{Babcock2015}
\textsc{Babcock, B.~A.} (2015): \enquote{Using Cumulative Prospect Theory to
  Explain Anomalous Crop Insurance Coverage Choice,} \emph{American Journal of
  Agricultural Economics}, 97, 1371--1384.

\bibitem[\protect\citeauthoryear{Barnett, Black, Hu, and Skees}{Barnett
  et~al.}{2005}]{BarnettBlackEtAl2005}
\textsc{Barnett, B.~J., J.~R. Black, Y.~Hu, and J.~R. Skees} (2005):
  \enquote{{Is Area Yield Insurance Competitive with Farm Yield Insurance?}}
  \emph{Journal of Agricultural and Resource Economics}, 30, 1--17.

\bibitem[\protect\citeauthoryear{Barnett and Mahul}{Barnett and
  Mahul}{2007}]{BarnettMahul2007}
\textsc{Barnett, B.~J. and O.~Mahul} (2007): \enquote{Weather Index Insurance
  for Agriculture and Rural Areas in Lower-Income Countries,} \emph{American
  Journal of Agricultural Economics}, 89, 1241--1247.

\bibitem[\protect\citeauthoryear{Barr\'e, Stoeffler, and Carter}{Barr\'e
  et~al.}{2016}]{BarreStoefflerEtAl2016}
\textsc{Barr\'e, T., Q.~Stoeffler, and M.~Carter} (2016): \enquote{Assessing
  index insurance: conceptual approach and empirical illustration from Burkina
  Faso,} Tech. rep., University of California Davis.

\bibitem[\protect\citeauthoryear{Binswanger-Mkhize}{Binswanger-Mkhize}{2012}]{Binswanger-Mkhize2012}
\textsc{Binswanger-Mkhize, H.~P.} (2012): \enquote{Is There Too Much Hype about
  Index-based Agricultural Insurance?} \emph{The Journal of Development
  Studies}, 48, 187--200.

\bibitem[\protect\citeauthoryear{Boryan, Yang, Mueller, and Craig}{Boryan
  et~al.}{2011}]{BoryanEtAl2011}
\textsc{Boryan, C., Z.~Yang, R.~Mueller, and M.~Craig} (2011):
  \enquote{Monitoring US agriculture: the US Department of Agriculture,
  National Agricultural Statistics Service, Cropland Data Layer Program,}
  \emph{Geocarto International}, 26, 341--358.

\bibitem[\protect\citeauthoryear{Boucher, Carter, and Guirkinger}{Boucher
  et~al.}{2008}]{BoucherCarterEtAl2008}
\textsc{Boucher, S., M.~Carter, and C.~Guirkinger} (2008): \enquote{Risk
  Rationing and Wealth Effects in Credit Markets: Implications for Agricultural
  Development,} \emph{American Journal of Agricultural Economics}, 90(2),
  409--423.

\bibitem[\protect\citeauthoryear{Bourgeon and Chambers}{Bourgeon and
  Chambers}{2003}]{BourgeonChambers2003}
\textsc{Bourgeon, J.-M. and R.~G. Chambers} (2003): \enquote{Optimal Area-Yield
  Crop Insurance Reconsidered,} \emph{American Journal of Agricultural
  Economics}, 85, 590--604.

\bibitem[\protect\citeauthoryear{Breustedt, Bokusheva, and
  Heidelbach}{Breustedt et~al.}{2008}]{BreustedtBokushevaEtAl2008}
\textsc{Breustedt, G., R.~Bokusheva, and O.~Heidelbach} (2008):
  \enquote{Evaluating the Potential of Index Insurance Schemes to Reduce Crop
  Yield Risk in an Arid Region,} \emph{Journal of Agricultural Economics}, 59,
  312--328.

\bibitem[\protect\citeauthoryear{Carriker, Williams, Barnaby, and
  Black}{Carriker et~al.}{1991}]{CarrikerWilliamsEtAl1991}
\textsc{Carriker, G.~L., J.~R. Williams, G.~A. Barnaby, and J.~R. Black}
  (1991): \enquote{Yield and Income Risk Reduction under Alternative Crop
  Insurance and Disaster Assistance Designs,} \emph{Western Journal of
  Agricultural Economics}, 16, 238--250.

\bibitem[\protect\citeauthoryear{Carter, de~Janvry, Sadoulet, and
  Sarris}{Carter et~al.}{2017}]{CarterJanvryEtAl2017}
\textsc{Carter, M., A.~de~Janvry, E.~Sadoulet, and A.~Sarris} (2017):
  \enquote{Index Insurance for Developing Country Agriculture: A Reassessment,}
  \emph{Annual Review of Resource Economics}, 9, 421--438.

\bibitem[\protect\citeauthoryear{Clarke}{Clarke}{2016}]{Clarke2016}
\textsc{Clarke, D.~J.} (2016): \enquote{A Theory of Rational Demand for Index
  Insurance,} \emph{American Economic Journal: Microeconomics}, 8, 283--306.

\bibitem[\protect\citeauthoryear{Cole, Gin\'e, Tobacman, Topalova, Townsend,
  and Vickery}{Cole et~al.}{2013}]{ColeGineEtAl2013}
\textsc{Cole, S., X.~Gin\'e, J.~Tobacman, P.~Topalova, R.~Townsend, and
  J.~Vickery} (2013): \enquote{Barriers to Household Risk Management: Evidence
  from India,} \emph{American Economic Journal: Applied Economics}, 5, 104--35.

\bibitem[\protect\citeauthoryear{Cole and Xiong}{Cole and
  Xiong}{2017}]{ColeXiong2017}
\textsc{Cole, S.~A. and W.~Xiong} (2017): \enquote{Agricultural Insurance and
  Economic Development,} \emph{Annual Review of Economics}, 9, 235--262.

\bibitem[\protect\citeauthoryear{Dado, Deines, and Lobell}{Dado
  et~al.}{2019}]{DadoDeinesEtAl2019}
\textsc{Dado, W., J.~M. Deines, and D.~B. Lobell} (2019): \enquote{Improving
  satellite-based soybean yield mapping across irrigated and rain-fed
  conditions,} in \emph{American Geophysical Union, Fall Meeting}.

\bibitem[\protect\citeauthoryear{Deines, Dado, Patel, and Lobell}{Deines
  et~al.}{2019{\natexlab{a}}}]{DeinesDadoEtAl2019}
\textsc{Deines, J.~M., W.~Dado, R.~Patel, and D.~B. Lobell}
  (2019{\natexlab{a}}): \enquote{Insights into Effective Satellite Crop Yield
  Estimation from an Extensive Ground Truth Dataset in the US Corn Belt,} in
  \emph{American Geophysical Union Fall Meeting}.

\bibitem[\protect\citeauthoryear{Deines, Wang, and Lobell}{Deines
  et~al.}{2019{\natexlab{b}}}]{DeinesWangEtAl2019}
\textsc{Deines, J.~M., S.~Wang, and D.~B. Lobell} (2019{\natexlab{b}}):
  \enquote{Satellites reveal a small positive yield effect from conservation
  tillage across the {US} Corn Belt,} \emph{Environmental Research Letters},
  14, 124038.

\bibitem[\protect\citeauthoryear{Deng, Barnett, Hoogenboom, Yu, and
  Garcia}{Deng et~al.}{2008}]{DengBarnettEtAl2008}
\textsc{Deng, X., B.~J. Barnett, G.~Hoogenboom, Y.~Yu, and A.~G.~y. Garcia}
  (2008): \enquote{{Alternative Crop Insurance Indexes},} \emph{Journal of
  Agricultural and Applied Economics}, 40, 223--237.

\bibitem[\protect\citeauthoryear{Deng, Barnett, and Vedenov}{Deng
  et~al.}{2007}]{DengBarnettEtAl2007}
\textsc{Deng, X., B.~J. Barnett, and D.~V. Vedenov} (2007): \enquote{Is There a
  Viable Market for Area-Based Crop Insurance?} \emph{American Journal of
  Agricultural Economics}, 89, 508--519.

\bibitem[\protect\citeauthoryear{Dercon}{Dercon}{1998}]{Dercon1998}
\textsc{Dercon, S.} (1998): \enquote{Wealth, risk and activity choice: cattle
  in Western Tanzania,} \emph{Journal of Development Economics}, 55, 1 -- 42.

\bibitem[\protect\citeauthoryear{Dercon}{Dercon}{2002}]{Dercon2002}
---\hspace{-.1pt}---\hspace{-.1pt}--- (2002): \enquote{Income Risk, Coping
  Strategies, and Safety Nets,} \emph{The World Bank Research Observer}, 17,
  141--166.

\bibitem[\protect\citeauthoryear{Du, Feng, and Hennessy}{Du
  et~al.}{2017}]{DuFengEtAl2017}
\textsc{Du, X., H.~Feng, and D.~A. Hennessy} (2017): \enquote{{Rationality of
  Choices in Subsidized Crop Insurance Markets},} \emph{American Journal of
  Agricultural Economics}, 99, 732--756.

\bibitem[\protect\citeauthoryear{Elabed, Bellemare, Carter, and
  Guirkinger}{Elabed et~al.}{2013}]{ElabedBellemareEtAl2013}
\textsc{Elabed, G., M.~F. Bellemare, M.~R. Carter, and C.~Guirkinger} (2013):
  \enquote{Managing basis risk with multiscale index insurance,}
  \emph{Agricultural Economics}, 44, 419--431.

\bibitem[\protect\citeauthoryear{Elabed and Carter}{Elabed and
  Carter}{2015}]{ElabedCarter2015}
\textsc{Elabed, G. and M.~R. Carter} (2015): \enquote{Compound-risk aversion,
  ambiguity and the willingness to pay for microinsurance,} \emph{Journal of
  Economic Behavior \& Organization}, 118, 150 -- 166, economic Experiments in
  Developing Countries.

\bibitem[\protect\citeauthoryear{Feng, Du, and Hennessy}{Feng
  et~al.}{2020}]{FengDuEtAl2020}
\textsc{Feng, H., X.~Du, and D.~A. Hennessy} (2020): \enquote{Depressed demand
  for crop insurance contracts, and a rationale based on third generation
  Prospect Theory,} \emph{Agricultural Economics}, 51, 59--73.

\bibitem[\protect\citeauthoryear{Flatnes, Carter, and Mercovich}{Flatnes
  et~al.}{2018}]{FlatnesCarterEtAl2018}
\textsc{Flatnes, J.~E., M.~R. Carter, and R.~Mercovich} (2018):
  \enquote{Improving the Quality of Index Insurance with a Satellite-based
  Conditional Audit Contract,} Tech. rep., Basis, Feed the Future Innovation
  Lab for Markets, Risk and Resilience, UC Davis.

\bibitem[\protect\citeauthoryear{Groemping}{Groemping}{2006}]{Groemping2006}
\textsc{Groemping, U.} (2006): \enquote{Relative Importance for Linear
  Regression in R: The Package relaimpo,} \emph{Journal of Statistical
  Software, Articles}, 17, 1--27.

\bibitem[\protect\citeauthoryear{Harri, Coble, Ker, and Goodwin}{Harri
  et~al.}{2011}]{HarriCobleEtAl2011}
\textsc{Harri, A., K.~H. Coble, A.~P. Ker, and B.~J. Goodwin} (2011):
  \enquote{Relaxing Heteroscedasticity Assumptions in Area-Yield Crop Insurance
  Rating,} \emph{American Journal of Agricultural Economics}, 93, 707--717.

\bibitem[\protect\citeauthoryear{Hennessy}{Hennessy}{2006}]{Hennessy2006}
\textsc{Hennessy, D.~A.} (2006): \enquote{On Monoculture and the Structure of
  Crop Rotations,} \emph{American Journal of Agricultural Economics}, 88, 900.

\bibitem[\protect\citeauthoryear{Jensen, Barrett, and Mude}{Jensen
  et~al.}{2016}]{JensenBarrettEtAl2016}
\textsc{Jensen, N.~D., C.~B. Barrett, and A.~G. Mude} (2016): \enquote{{Index
  Insurance Quality and Basis Risk: Evidence from Northern Kenya},}
  \emph{American Journal of Agricultural Economics}, 98, 1450--1469.

\bibitem[\protect\citeauthoryear{Jensen, Mude, and Barrett}{Jensen
  et~al.}{2018}]{JensenMudeEtAl2018}
\textsc{Jensen, N.~D., A.~G. Mude, and C.~B. Barrett} (2018): \enquote{How
  basis risk and spatiotemporal adverse selection influence demand for index
  insurance: Evidence from northern Kenya,} \emph{Food Policy}, 74, 172 -- 198.

\bibitem[\protect\citeauthoryear{Jin, Azzari, and Lobell}{Jin
  et~al.}{2017}]{JinAzzariEtAl2017}
\textsc{Jin, Z., G.~Azzari, and D.~B. Lobell} (2017): \enquote{Improving the
  accuracy of satellite-based high-resolution yield estimation: A test of
  multiple scalable approaches,} \emph{Agricultural and Forest Meteorology},
  247, 207 -- 220.

\bibitem[\protect\citeauthoryear{Just, Calvin, and Quiggin}{Just
  et~al.}{1999}]{JustCalvinEtAl1999}
\textsc{Just, R.~E., L.~Calvin, and J.~Quiggin} (1999): \enquote{Adverse
  Selection in Crop Insurance: Actuarial and Asymmetric Information
  Incentives,} \emph{American Journal of Agricultural Economics}, 81, 834--849.

\bibitem[\protect\citeauthoryear{Karlan, Osei, Osei-Akoto, and Udry}{Karlan
  et~al.}{2014}]{KarlanOseiEtAl2014}
\textsc{Karlan, D., R.~Osei, I.~Osei-Akoto, and C.~Udry} (2014):
  \enquote{Agricultural Decisions after Relaxing Credit and Risk Constraints,}
  \emph{The Quarterly Journal of Economics}, 129, 597--652.

\bibitem[\protect\citeauthoryear{Lobell and Azzari}{Lobell and
  Azzari}{2017}]{LobellAzzari2017}
\textsc{Lobell, D.~B. and G.~Azzari} (2017): \enquote{Satellite detection of
  rising maize yield heterogeneity in the U.S. Midwest,} \emph{Environmental
  Research Letters}, 12, 014014.

\bibitem[\protect\citeauthoryear{Lobell, Thau, Seifert, Engle, and
  Little}{Lobell et~al.}{2015}]{LobellEtAl2015}
\textsc{Lobell, D.~B., D.~Thau, C.~Seifert, E.~Engle, and B.~Little} (2015):
  \enquote{A scalable satellite-based crop yield mapper,} \emph{Remote Sensing
  of Environment}, 164, 324 -- 333.

\bibitem[\protect\citeauthoryear{Mahul}{Mahul}{1999}]{Mahul1999}
\textsc{Mahul, O.} (1999): \enquote{{Optimum Area Yield Crop Insurance},}
  \emph{American Journal of Agricultural Economics}, 81, 75--82.

\bibitem[\protect\citeauthoryear{Miranda and Farrin}{Miranda and
  Farrin}{2012}]{MirandaFarrin2012}
\textsc{Miranda, M. and K.~Farrin} (2012): \enquote{Index Insurance for
  Developing Countries,} \emph{Applied Economic Perspectives and Policy}, 34,
  391--427.

\bibitem[\protect\citeauthoryear{Miranda}{Miranda}{1991}]{Miranda1991}
\textsc{Miranda, M.~J.} (1991): \enquote{Area-Yield Crop Insurance
  Reconsidered,} \emph{American Journal of Agricultural Economics}, 73,
  233--242.

\bibitem[\protect\citeauthoryear{Schnitkey, Coppess, Paulson, and
  Zulauf}{Schnitkey et~al.}{2015}]{SchnitkeyCoppessEtAl2015}
\textsc{Schnitkey, G., J.~Coppess, N.~Paulson, and C.~Zulauf} (2015):
  \enquote{Perspectives on Commodity Program Choices Under the 2014 Farm Bill,}
  \emph{Farmdoc Daily}, 5:111.

\bibitem[\protect\citeauthoryear{Seifert, Azzari, and Lobell}{Seifert
  et~al.}{2018}]{SeifertAzzariEtAl2018}
\textsc{Seifert, C.~A., G.~Azzari, and D.~B. Lobell} (2018): \enquote{Satellite
  detection of cover crops and their effects on crop yield in the Midwestern
  United States,} \emph{Environmental Research Letters}, 13, 064033.

\bibitem[\protect\citeauthoryear{Seifert, Roberts, and Lobell.}{Seifert
  et~al.}{2017}]{SeifertEtAl2017}
\textsc{Seifert, C.~A., M.~J. Roberts, and D.~B. Lobell.} (2017):
  \enquote{Continuous Corn and Soybean Yield Penalties across Hundreds of
  Thousands of Fields,} \emph{Agronomy Journal}, 109, 541--548.

\bibitem[\protect\citeauthoryear{Skees, Black, and Barnett}{Skees
  et~al.}{1997}]{SkeesBlackEtAl1997}
\textsc{Skees, J.~R., J.~R. Black, and B.~J. Barnett} (1997):
  \enquote{Designing and Rating an Area Yield Crop Insurance Contract,}
  \emph{American Journal of Agricultural Economics}, 79, 430--438.

\bibitem[\protect\citeauthoryear{Smith, Chouinard, and Baquet}{Smith
  et~al.}{1994}]{SmithChouinardEtAl1994}
\textsc{Smith, V.~H., H.~H. Chouinard, and A.~E. Baquet} (1994):
  \enquote{{Almost Ideal Area Yield Crop Insurance Contracts},}
  \emph{Agricultural and Resource Economics Review}, 23, 1--9.

\bibitem[\protect\citeauthoryear{Stigler}{Stigler}{2018}]{Stigler2018}
\textsc{Stigler, M.} (2018): \enquote{Supply response at the field-level:
  disentangling area and yield effects,} Tech. rep., UC Davis, ARE.

\bibitem[\protect\citeauthoryear{Stigler}{Stigler}{2019{\natexlab{a}}}]{Stigler_rotation_2019}
---\hspace{-.1pt}---\hspace{-.1pt}--- (2019{\natexlab{a}}): \enquote{Measuring
  rotation effects in the US Corn Belt,} Tech. rep., Chapter 2 of dissertation,
  \url{https://github.com/MatthieuStigler/MatthieuStigler.github.io/raw/master/docs/rotation_effects_Stigler_standalone.pdf}.

\bibitem[\protect\citeauthoryear{Stigler}{Stigler}{2019{\natexlab{b}}}]{Stigler_stlyised_CB}
---\hspace{-.1pt}---\hspace{-.1pt}--- (2019{\natexlab{b}}): \enquote{US Corn
  Belt: a satellite view,} Tech. rep., Chapter 1 of PhD dissertation,
  \url{https://github.com/MatthieuStigler/MatthieuStigler.github.io/raw/master/docs/Chapter1\_Stylised\_facts\_standalone.pdf}.

\bibitem[\protect\citeauthoryear{Tversky and Kahneman}{Tversky and
  Kahneman}{1992}]{TverskyKahneman1992}
\textsc{Tversky, A. and D.~Kahneman} (1992): \enquote{{Advances in Prospect
  Theory: Cumulative Representation of Uncertainty},} \emph{Journal of Risk and
  Uncertainty}, 5, 297--323.

\bibitem[\protect\citeauthoryear{Vercammen}{Vercammen}{2000}]{Vercammen2000}
\textsc{Vercammen, J.~A.} (2000): \enquote{{Constrained Efficient Contracts for
  Area Yield Crop Insurance},} \emph{American Journal of Agricultural
  Economics}, 82, 856--864.

\bibitem[\protect\citeauthoryear{Wang, Hanson, Myers, and Black}{Wang
  et~al.}{1998}]{WangHansonEtAl1998}
\textsc{Wang, H.~H., S.~D. Hanson, R.~J. Myers, and J.~R. Black} (1998):
  \enquote{The Effects of Crop Yield Insurance Designs on Farmer Participation
  and Welfare,} \emph{American Journal of Agricultural Economics}, 80,
  806--820.

\bibitem[\protect\citeauthoryear{Wang, Di~Tommaso, Deines, and Lobell}{Wang
  et~al.}{2020}]{WangDiTommasoEtAl2020}
\textsc{Wang, S., S.~Di~Tommaso, J.~M. Deines, and D.~Lobell} (2020):
  \enquote{Mapping twenty years of corn and soybean across the US Midwest using
  the Landsat archive,} \emph{Scientific Data}, 7, 307.

\bibitem[\protect\citeauthoryear{Ye, Hu, Barnett, Wang, and Gao}{Ye
  et~al.}{2020}]{YeHuEtAl2020}
\textsc{Ye, T., W.~Hu, B.~J. Barnett, J.~Wang, and Y.~Gao} (2020):
  \enquote{Area Yield Index Insurance or Farm Yield Crop Insurance? Chinese
  Perspectives on Farmers' Welfare and Government Subsidy Effectiveness,}
  \emph{Journal of Agricultural Economics}, 71, 144--164.

\end{thebibliography}

\end{document}